\documentclass[review]{elsarticle}

\usepackage{lineno,hyperref}
\modulolinenumbers[5]

\usepackage{amsmath,amsthm,amssymb,epsfig,bm,amsmath}
 \usepackage{graphicx}
\usepackage{float}
\usepackage{amsmath}
\usepackage{mhchem}
\usepackage{xparse}
\usepackage{adjustbox}
\usepackage{MnSymbol}% http://ctan.org/pkg/mnsymbol
\usepackage{mathdots}% http://ctan.org/pkg/mathdots
\usepackage{stmaryrd, isomath, wasysym, amsxtra}
\usepackage{makecell}
\usepackage{listings}
\usepackage[printwatermark]{xwatermark}
\usepackage{footnote}
\usepackage[flushleft]{threeparttable}
\usepackage{multirow}
\usepackage{relsize}
\usepackage{lettrine}
\usepackage{courier}
\usepackage{color} %red, green, blue, yellow, cyan, magenta, black, white
\definecolor{mygreen}{RGB}{28,172,0} % color values Red, Green, Blue
\definecolor{mylilas}{RGB}{170,55,241}
\usepackage{amsmath,amsthm,amssymb,amsfonts} % this for /equation* (without numbering)
\usepackage{graphicx} % for figures
\usepackage[margin=1in]{geometry} % changing the margins
\usepackage[yyyymmdd]{datetime}
\usepackage{algorithm}
\usepackage[noend]{algpseudocode}
\usepackage{lipsum}
\usepackage{setspace}
\usepackage{siunitx}
\usepackage{pdfpages}
\usepackage{booktabs,caption}
\usepackage[]{hyperref}
\hypersetup{
  colorlinks   = true, %Colours links instead of ugly boxes
  urlcolor     = blue, %Colour for external hyperlinks
  linkcolor    = black, %Colour of internal links
  citecolor   = blue %Colour of citations
}
\usepackage[utf8]{inputenc}
 
\usepackage{listings}
\usepackage{color}
\definecolor{codegreen}{rgb}{0,0.6,0}
\definecolor{codegray}{rgb}{0.5,0.5,0.5}
\definecolor{codepurple}{rgb}{0.58,0,0.82}
\definecolor{backcolour}{rgb}{0.95,0.95,0.92}
\lstdefinestyle{mystyle}{
    backgroundcolor=\color{backcolour},   
    commentstyle=\color{codegreen},
    keywordstyle=\color{magenta},
    numberstyle=\tiny\color{codegray},
    stringstyle=\color{codepurple},
    basicstyle=\footnotesize,
    breakatwhitespace=false,         
    breaklines=true,                 
    captionpos=b,                    
    keepspaces=true,                 
    numbers=left,                    
    numbersep=5pt,                  
    showspaces=false,                
    showstringspaces=false,
    showtabs=false,                  
    tabsize=2,
    escapeinside={<@}{@>},
}
 
\usepackage{booktabs} % For \toprule, \midrule and \bottomrule
\usepackage{siunitx} % Formats the units and values

\usepackage{mathtools} % for arrows
\usepackage{gensymb} % To write degree C 
\usepackage{mhchem} % To write chemical symbols
\usepackage{fixltx2e} % To write subscripts and superscripts 
\usepackage{bbm}

\usepackage{multirow}

\usepackage{fancyhdr} % customize the footer and header
\theoremstyle{definition}

\theoremstyle{definition}

\theoremstyle{remark}

\usepackage{soul} %To highlight the text 
\soulregister\cite7
\soulregister\ref7
\soulregister\pageref7
\usepackage{bm}

\usepackage{framed} % Framing content

\usepackage{multicol} % Multiple columns environment

\usepackage{nomencl} % Nomenclature package
\usepackage{tikz}
\makenomenclature
\usepackage[resetlabels,labeled]{multibib}

\renewcommand*\nompreamble{\begin{multicols}{2}}

\renewcommand*\nompostamble{\end{multicols}}

\usepackage{amssymb}% http://ctan.org/pkg/amssymb
\usepackage{pifont}% http://ctan.org/pkg/pifont
\definecolor{light-gray}{gray}{0.95}

\DeclareUnicodeCharacter{2212}{-}

\journal{a Journal for Review}

\begin{document}

\begin{frontmatter}

\title{\large Fault Prognosis in Particle Accelerator Power Electronics Using Ensemble Learning}

\author{Majdi I. Radaideh$^{a*}$, Chris Pappas$^b$, Mark Wezensky$^c$, Pradeep Ramuhalli$^d$, Sarah Cousineau$^e$}

\cortext[mycorrespondingauthor]{Corresponding Author: Majdi I. Radaideh (radaidehmi@ornl.gov)}

\address{$^a$Machine Learning Engineer for Control of Complex Systems, Spallation Neutron Source, Oak Ridge National Laboratory, Oak Ridge, Tennessee 37830, United States}
\address{$^b$Power Electronics Engineer, Spallation Neutron Source, Oak Ridge National Laboratory, Oak Ridge, Tennessee 37830, United States}
\address{$^c$Control Systems Engineer, Spallation Neutron Source, Oak Ridge National Laboratory, Oak Ridge, Tennessee 37830, United States}
\address{$^d$Group Leader for Modern Nuclear Instrumentation and Control, Nuclear Energy and Fuel Cycle Division, Oak Ridge National Laboratory, Oak Ridge, Tennessee 37830, United States}
\address{$^e$Section Head for Accelerator Science and Technology, Spallation Neutron Source, Oak Ridge National Laboratory, Oak Ridge, Tennessee 37830, United States}

\begin{abstract}

\small

Early fault detection and fault prognosis are crucial to ensure efficient and safe operations of complex engineering systems such as the Spallation Neutron Source (SNS) and its power electronics (high voltage converter modulators). Following an advanced experimental facility setup that mimics SNS operating conditions, the authors successfully conducted 21 fault prognosis experiments, where fault precursors are introduced in the system to a degree enough to cause degradation in the waveform signals, but not enough to reach a real fault. Nine different machine learning techniques based on ensemble trees, convolutional neural networks, support vector machines, and hierarchical voting ensembles are proposed to detect the fault precursors. Although all 9 models have shown a perfect and identical performance during the training and testing phase, the performance of most models has decreased in the prognosis phase once they got exposed to real-world data from the 21 experiments. The hierarchical voting ensemble, which features multiple layers of diverse models, maintains a distinguished performance in early detection of the fault precursors with 95\% success rate (20/21 tests), followed by adaboost and extremely randomized trees with 52\% and 48\% success rates, respectively. The support vector machine models were the worst with only 24\% success rate (5/21 tests). The study concluded that a successful implementation of machine learning in the SNS or particle accelerator power systems would require a major upgrade in the controller and the data acquisition system to facilitate streaming and handling big data for the machine learning models. In addition, this study shows that the best performing models were diverse and based on the ensemble concept to reduce the bias and hyperparameter sensitivity of individual models.

\end{abstract}


\end{frontmatter}

%\linenumbers

% \pagenumbering{gobble}
\setstretch{1.5}

\section{Introduction}

Early fault detection and fault prognosis are crucial to ensure efficient and safe operations of complex engineering systems. Fault prognosis can detect and predict faults early in time before the faults cause a system damage, which can be done by examining fault symptoms as early as possible \cite{vachtsevanos2006intelligent}. From now on, we use ``fault precursors'' to refer to fault symptoms or fault indicator signals. Advancing prognosis approaches is vital to the success of the predictive maintenance, which is unlike reactive or preventive maintenance, relies on early fault detection \cite{zhang2019data}. Predictive maintenance has several advantages over preventive maintenance, which include (1) improving machine availability, (2) extension of the machine operation life, (3) prevention of catastrophic failures, and (4) optimizing the resources for preventive maintenance \cite{fernandes2022machine}. All these reasons can lead to improved productivity by the machine. 

Fault prognosis and predictive maintenance with machine learning techniques have already illustrated a promising potential. The study by \cite{arunthavanathan2021deep} used convolutional neural networks (CNN) and long short-term memory (LSTM) to forecast the system parameters and an unsupervised one-class support vector machine for fault precursor detection. The approach was assessed using the Tennessee Eastman process fault data. Similarly, a deep learning approach (a combination of sparse autoencoder and fully-connected layers) was proposed by \cite{luo2018early} for early fault detection by automatically selecting the impulse responses from vibration signals in a time-varying system. In another study \cite{shao2017enhancement}, a deep autoencoder was developed for rotating machinery fault diagnosis, which consists of denoising autoencoder and contractive autoencoder layers to enhance the feature extraction ability. The authors of \cite{wang2016wind} have developed deep neural network framework for monitoring the health of wind turbine gearboxes based on the lubricant pressure data, which shows that the deep learning model is more capable than other classical machine learning methods (e.g. k-nearest neighbours, support vector machine). A predictive maintenance model based on LSTM and generative adversarial networks (GAN) was proposed by \cite{liu2021novel} to determine the state of the machine and the fault in advance. Ensemble or tree-based machine learning methods also showed promise in fault prognosis applications such as Adaboost \cite{kozjek2017knowledge}, isolation forests \cite{kolokas2020generic}, random forests \cite{syafrudin2018performance}, random survival forests \cite{bukkapatnam2019machine}, and gradient boosting trees \cite{zhang2020cpps}. Comprehensive surveys of machine learning methods for fault prognosis were conducted by \cite{zhang2019review,fernandes2022machine}.

In fault detection and classification, machine learning and neural networks have also progressed as described in this comprehensive survey \cite{mohd2020neural}. Fault detection applications of neural networks have been demonstrated in different energy and electronics fields such as integrated energy systems \cite{wang2021fault}, large-scale power systems with LSTM \cite{belagoune2021deep}, fusion energy devices (Tokamaks) \cite{mohapatra2020real}, photovoltaic systems \cite{hajji2021multivariate}, building energy consumption \cite{bode2020real}, and similar others. Similar fault detection efforts with non-neural network methods were demonstrated by \cite{taqvi2020fault} using nonlinear autoregressive with exogenous input and by \cite{agasthian2019fault} using support vector machine optimized by the Cuckoo search algorithm.  

Particle accelerators, such as the spallation neutron source (SNS) \cite{henderson2014spallation} and CERN, are complex engineering systems that use electromagnetic fields to propel charged particles to very high speeds and energies to use them for fundamental research applications. The interest in machine learning for control applications in particle accelerators can be seen in these studies \cite{nguyen1991accelerator, edelen2016neural}. Uncertainty-aware anomaly detection framework of the errant beam pulses was developed by \cite{blokland2021uncertainty} using Siamese neural networks with ResNet blocks. For a beam-based study with real measured data, the authors of \cite{rescic2020predicting} employed different machine learning binary classifiers (e.g. logistic regression, gradient boosting, random forests) to predict system failure. The results demonstrated a promising performance with failure prediction accuracy up to 92\% after fine tuning the classifier hyperparameters. The fault detection effort was then improved in a subsequent study by the team for pre-emptive detection of machine trips in the SNS \cite{revsvcivc2022improvements}. Lastly, a recent study by \cite{felsberger2020explainable} investigates a deep learning model based on CNN for fault prognosis in a particle accelerator system (CERN). Despite promising performance, the authors found difficulties in predicting certain failures due to the lack of data, given the authors have relied on historical data. A similar effort for using adaptive neural networks for time-varying beam control was demonstrated in \cite{scheinker2021adaptive}.

%In addition, anomaly detection was applied to an HVCM module using discrete cosine transform \cite{pappas2021machine}, showing good results given the limited data. In this work, Recurrent AutoEncoders (RAE) will be applied, given their promising performance in other domains and after collecting more data from the HVCM system.
%The HVCMs have been a significant source of lost user time at the SNS, as is shown by \cite{radaideh2022time}. A previous study on using discrete cosine transform for anomaly detection in HVCM waveforms was done by \cite{pappas2021machine}, while \cite{radaideh2022real} developed advanced recurrent neural network autoencoder models for time series anomaly detection in the HVCMs powering the RFQ. Further efforts on applications of machine learning for fault detection in particle accelerators include application of of variety of binary classifiers \cite{rescic2020predicting}, Siamese neural networks \cite{blokland2021uncertainty}, adaptive neural networks for time-varying beam control \cite{scheinker2021adaptive}, and similar others \cite{edelen2016neural}.

In this work, we explore machine learning methods for fault prognosis in the power systems of the SNS, called high voltage converter modulators (HVCM). HVCMs are used in the SNS to power the klystrons that accelerate the charged particles to about 90\% of the speed of light. HVCMs continue to be problematic systems for the SNS, that experience a wide range of failures from mild to catastrophic, causing reliability issues and lost beam time for the SNS. HVCMs are ranked among the top sources of downtime in the SNS \cite{radaideh2022time}. Compared to our previous study \cite{radaideh2022time}, which focused on instantaneous anomaly detection in HVCM signals using recurrent neural networks, this work extends the infrastructure and the methodology to allow fault prognosis capabilities in the HVCM, which enable operator intervention and predictive maintenance to be performed in most of the fault scenarios. The previous study \cite{radaideh2022time} was limited to a very short time scale that allows a graceful shutdown of the facility, due to limitations in the controller and the data acquisition system, all of which are resolved in this work. The major goal of this work is to develop and demonstrate a test facility that shows fault prognosis capabilities that warn the HVCM control room of impending failures or long term degradation of components. For accurate fault prognosis, the authors relied on three major components that highlight the main accomplishments of this work: (1) improved data acquisition system, (2) fast and continuous data streaming, and (3) improved fault prognosis models. To accomplish these goals, an advanced experimental setup that simulates SNS operating conditions and a collection of fault test scenarios are prepared and used to test our proposed fault prognosis and detection methods based on machine learning.

The remaining sections of this work are organized as follows: Section \ref{sec:exp} describes the experimental setup which involves a radio-frequency test facility established for data streaming, model development, and model testing. Section \ref{sec:method} highlights the methodology implemented in this work, which includes data preparation, machine learning models, and performance metrics. The results of this work are presented and discussed in section \ref{sec:res}, followed by the conclusions of this work in section \ref{sec:conc}. 

\section{Experimental Setup}
\label{sec:exp}

The Spallation Neutron Source (SNS) at Oak Ridge National Laboratory (ORNL) accelerates protons to high speeds, which are used to produce neutron beams for neutron scattering and materials research \cite{henderson2014spallation}. The beam is accelerated in a linear accelerator consisting of a Radio Frequency Quadrupole (RFQ) section, a Drift Tube Linac (DTL) section, a Couple Cavity Linac (CCL) section, and a Superconducting Linac (SCL) section. The accelerating cavities in each of these sections are fed by high power microwave amplifiers or klystrons. The klystrons are powered by High Voltage Converter Modulators (HVCM). The HVCMs can drive as many as 10 klystrons, depending on the klystron type and which section of the linac the klystron is located. There are a total of 15 HVCMs in the SNS, driving a total of 92 klystrons, where the HVCM powering the RFQ section (3 klystrons) was the subject of the analysis in our previous effort \cite{radaideh2022time}. We recently shared the normal and fault data collected from the 15 HVCMs of the SNS \cite{radaideh2022real}, collected over 2 years with the data being sparse in time (recorded signals can be separated by hours and even days). Given their time sparsity, that data \cite{radaideh2022real} or models can be used for fault classification and identification but not for fault prognosis, which is the motivation behind this work.  

Given the complexity of the SNS structure, it is very difficult to apply the proposed methods of this work directly on the SNS HVCMs since they would cause interruptions to the daily operation of the facility, especially that our experiments in this work involve fault induction tests, which can cause serious problems to the SNS. Alternatively, the radio-frequency test facility (RFTF) at the SNS provides a robust option to test new methods, components, or materials in an environment similar to the SNS environment. It is a known practice in the particle accelerator community to test new concepts in a test environment before applying it to the main accelerator, given that changing something in the SNS directly requires many approvals and careful investigation. \textit{Therefore, this section and the rest of this work focus on the RFTF setup.} 

\subsection{RFTF HVCM Description}

\begin{figure*}[!h]
   \centering
   \includegraphics*[width=\textwidth]{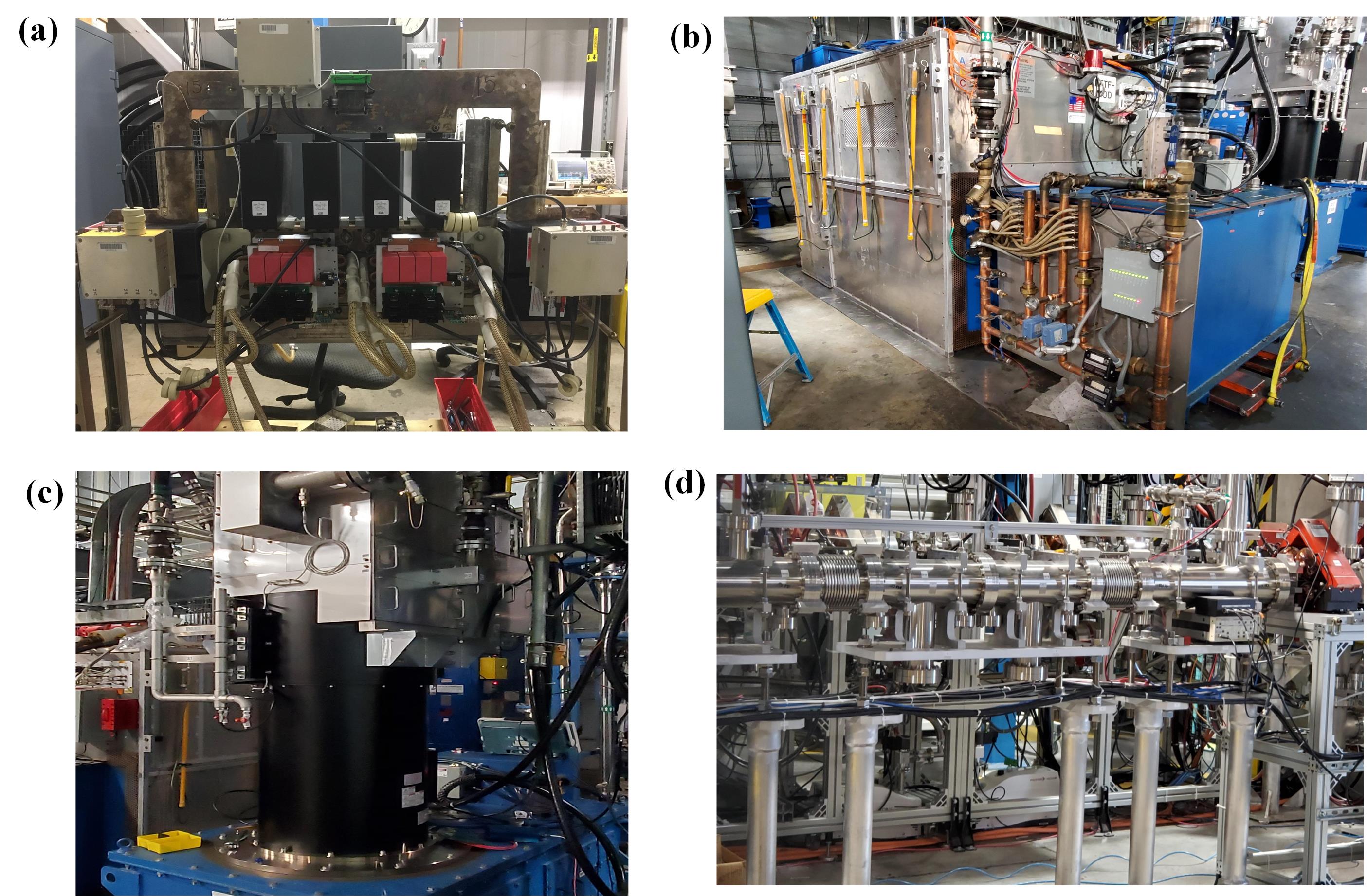}
   \caption{Components from the RFTF experimental setup: (a) HVCM H-bridge circuit, (b) HVCM insulation tank, (c) klystron (linear-beam vacuum tube), (d) section of the beam line}
   \label{fig:hvcm_rico}
 \end{figure*}

The authors have used the RFTF facility at the SNS in this work to develop and test machine learning models for fault prognosis. Figure \ref{fig:hvcm_rico} shows some of the major components from the RFTF facility. Figure \ref{fig:hvcm_rico}(a) shows the H-bridge switch plate of the HVCM while (b) shows the high voltage enclosure and insulation tank, which houses the HVCM assembly. Figure \ref{fig:hvcm_rico}(c) shows the linear-beam vacuum tube or the klystron, which is being powered by the HVCM. Lastly, Figure \ref{fig:hvcm_rico}(d) demonstrates a section of the beam line at which the particles can be accelerated (which can be powered by similar sources as the RFTF). The reader should notice here that the beam line is not directly utilized in this work and only shown for completeness of the facility description. 

\begin{figure*}[!h]
   \centering
   \includegraphics*[width=0.9\textwidth]{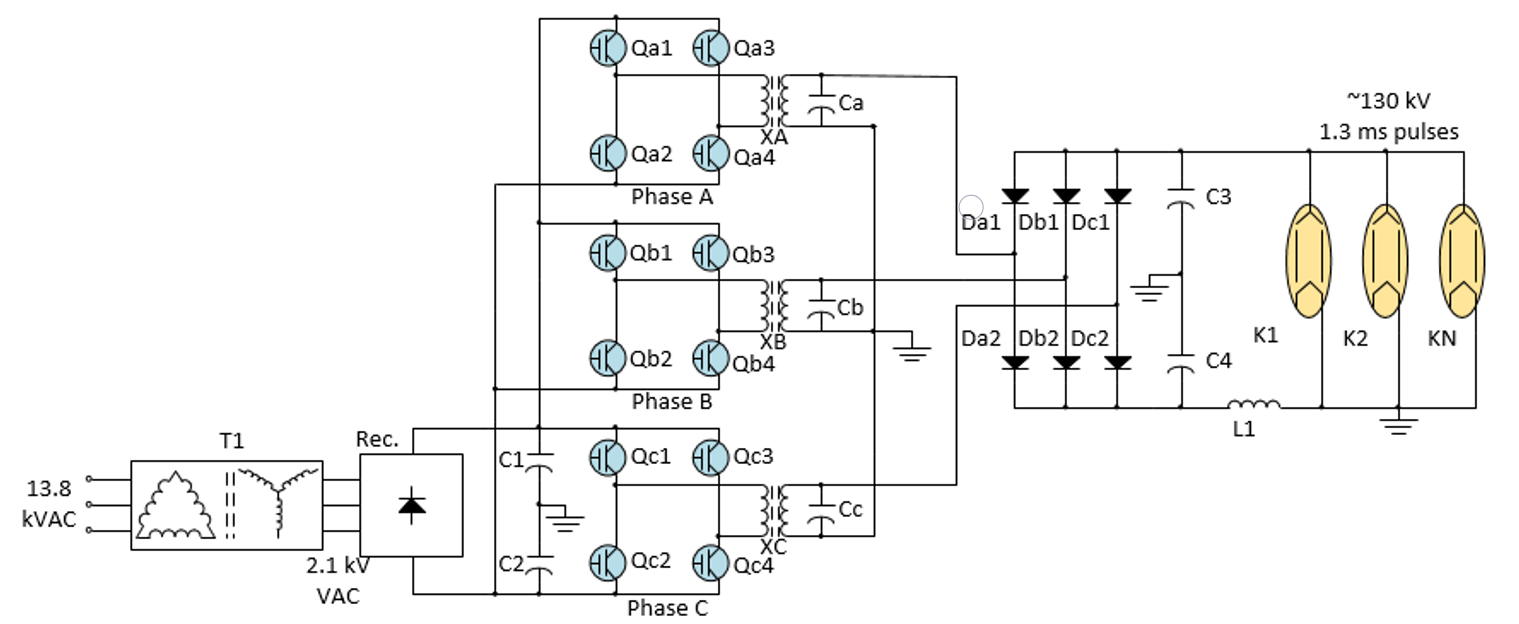}
   \caption{Simplified schematic of the HVCM circuit}
   \label{fig:hvcm}
\end{figure*}

Digging deeper into the HVCM structure, the HVCM circuit is shown in Figure \ref{fig:hvcm}, which can be summarised by the following events that occur in the system in a frequency of 60 Hz with a high voltage pulse width of 1.3 ms:
\begin{enumerate}
    \item  An input of 13.8 kVAC three-phase line power is converted to $\pm$1300 VDC by the transformer T1 (see Figure \ref{fig:hvcm}) and a six-pulse controlled rectifier circuit. Capacitors C1 and C2 in Figure \ref{fig:hvcm} filter this voltage and store sufficient charge to produce 1.3 ms pulses without excessive droop.
    \item The DC voltage is supplied to three-phase insulated-gate bipolar transistor (IGBT) H-bridge circuits, see Figure \ref{fig:hvcm_rico}(a), operating at a nominal switching frequency of 20 kHz. The IGBT switches are represented by Qa1-Qa4, Qb1-Qb4, and Qc1-Qc4 in Figure \ref{fig:hvcm}. The pulse transformers are used to step up the high power pulses to high voltage signals.
    \item The leakage inductance of the pulse transformers (XA, XB, XC) form a resonant circuit with the resonant capacitors (Ca, Cb, Cc) in Figure \ref{fig:hvcm}, giving the circuit a frequency dependent gain. 
    \item The high voltage bipolar pulses from the resonant capacitors are recombined and rectified by the diodes Da1 to Dc2. 
    \item The output pulses from the diodes, which have an apparent switching frequency of 120 kHz, are filtered by C3, C4 and L1, and applied to the cathode of the klystron in Figure \ref{fig:hvcm_rico}(c).
\end{enumerate}

\subsection{Data Streaming}
\label{sec:stream}
 
The HVCM in the RFTF uses PXI-based controller to (1) control the IGBT gating timing, (2) to ensure signal values of the pulse transformers remain in a safe range, (3) to set warning and trip levels for a variety of signals, and (4) to communicate with the control room and other auxiliary systems (e.g. personnel protection systems). More importantly the controller helps digitizing and save waveforms, which are the main source of data in this study. 

We performed multiple changes in the RFTF to be able to stream and save data continuously for machine learning efforts, compared to the sparse data collected in our previous study \cite{radaideh2022time}. First, the normal/fault files archived for the SNS main HVCMs require storage of approximately 30 MB of data when decimated to 2.5 MS/s. However, not all of this data is useful. Therefore, to reduce the massive amount of disk space required to store streamed data, the number of streamed waveform channels was reduced from 32 to 12 in the RFTF. Also, the record length was reduced from 3.6 ms to 1.5 ms with a sampling rate of 400 ns. These changes have reduced the size of each waveform file from 30 MB to approximately 540 kB, which facilitate data pre-processing for the machine learning models and significantly reduces data size. The data are saved to an external hard drive for the authors to use. The 12 waveforms recorded by the controller are (by referring to Figure \ref{fig:hvcm}):

\begin{enumerate}
    \item A+IGBT-I: The current passing through the IGBT switches (Qa1, Qa4) of phase A+ (unit: Ampere).
	\item A+*IGBT-I: The current passing through the IGBT switches (Qa2, Qa3) of phase A+* (unit: Ampere).
	\item B+IGBT-I: The current passing through the IGBT switches (Qb1, Qb4) of phase B+ (unit: Ampere).
	\item B+*IGBT-I: The current passing through the IGBT switches (Qb2, Qb3) of phase B+* (unit: Ampere).
	\item C+IGBT-I: The current passing through the IGBT switches (Qc1, Qc4) of phase C+ (unit: Ampere).
	\item C+*IGBT-I: The current passing through the IGBT switches (Qc2, Qc3) of phase C+* (unit: Ampere).
	\item Mod-I: Modulator current (unit: Ampere).
	\item A-Flux: Magnetic flux density for phase A transformer (unit: scaled).
	\item B-Flux: Magnetic flux density for phase B transformer (unit: scaled).
	\item C-Flux: Magnetic flux density for phase C transformer (unit: scaled).
	\item Mod-V: Modulator voltage (unit: kV).
	\item CB-V: Cap bank voltage (unit: V).
\end{enumerate}

The second improvement includes the ability to track waveform files when the HVCM is in the tuning mode. Tuning the HVCMs is normally done after maintenance on a particular HVCM. Tuning is done manually by experienced technicians and involves setting start and stop frequencies for IGBT gating to minimize droop, varying the start timing of the initial gate signals to minimize the likelihood of saturating the magnetic transformers. Tuning involves making incremental changes at reduced power while monitoring multiple signals such as klystron voltage, IGBT currents, and core magnetic flux to ensure they meet pulse requirements and remain within predetermined safe values. The data acquisition system records most of the waveforms during the tuning phase since the system would experience many changes.  

The third improvement involves the file saving rate during operation, which is no longer fixed by the controller, but determined by the user. For example, the user can record waveforms at a rate of a waveform file every 3 seconds when a large demand for data acquisition is present. The rate can be decreased to a file every 10 minutes when the streamed data are not needed.    
	
\begin{figure*}[!h]
   \centering
   \includegraphics*[width=0.7\textwidth]{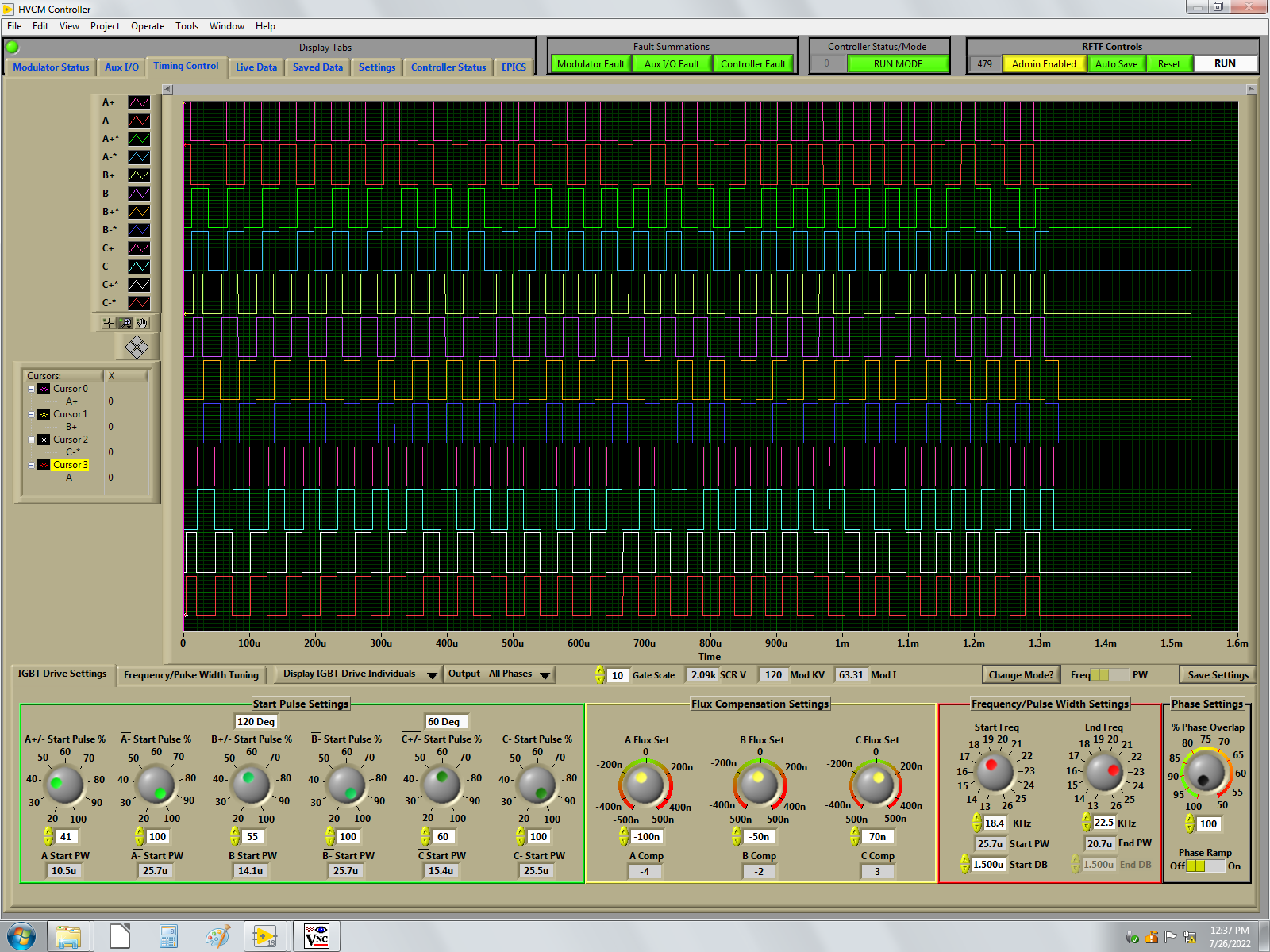}
   \caption{Screenshot of the RFTF HVCM live controller showing the settings and a timing diagram of the IGBT gate pulses}
   \label{fig:controller1}
\end{figure*}

\begin{figure*}[!h]
   \centering
   \includegraphics*[width=0.7\textwidth]{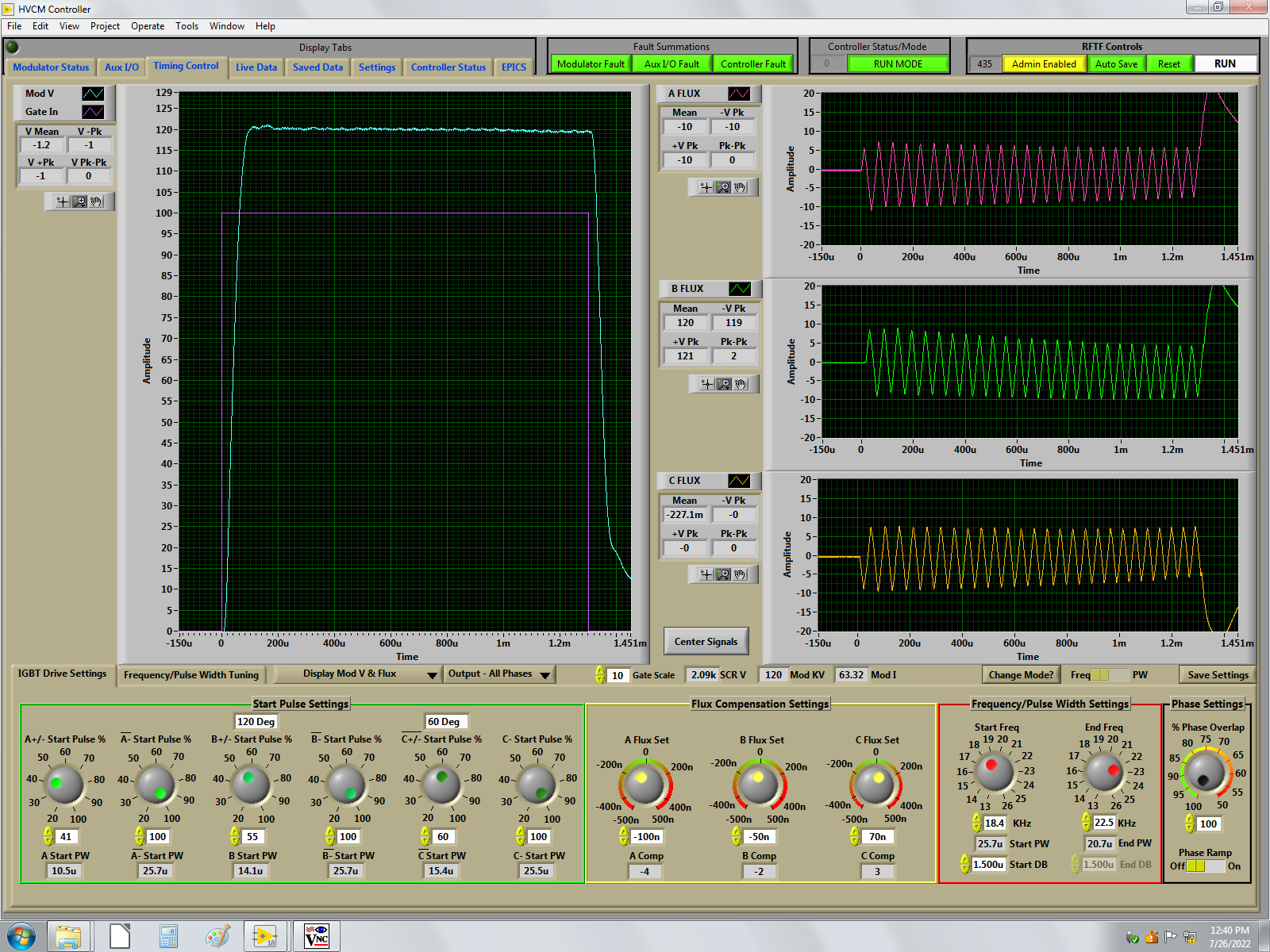}
   \caption{Screenshot of the RFTF HVCM live controller showing the settings, Mod-V diagram (left), and magnetic flux diagrams (A-Flux, B-Flux, C-Flux) on the right}
   \label{fig:controller2}
\end{figure*}

Figures \ref{fig:controller1}-\ref{fig:controller2} show live screenshots of the controller screen, which show the setting knobs that are used to tune the HVCM during startup. Also, plots of different waveforms are shown in the screen for the operator, which include a timing diagram of the IGBT gate pulses in Figure \ref{fig:controller1}. In the next section, we will describe how we utilize this data acquisition system and the setting knobs to model fault scenarios for prognosis purposes. 

%Due to the complexity of the experimental setup and to keep the scope of this work concise, more details about the experimental work are provided in a companion data article (pending the acceptance of this work). \textbf{The data article is attached as a supplementary material with this submission.}  

\section{Methodology}
\label{sec:method}

\begin{figure*}[!h]
   \centering
   \includegraphics*[width=0.8\textwidth]{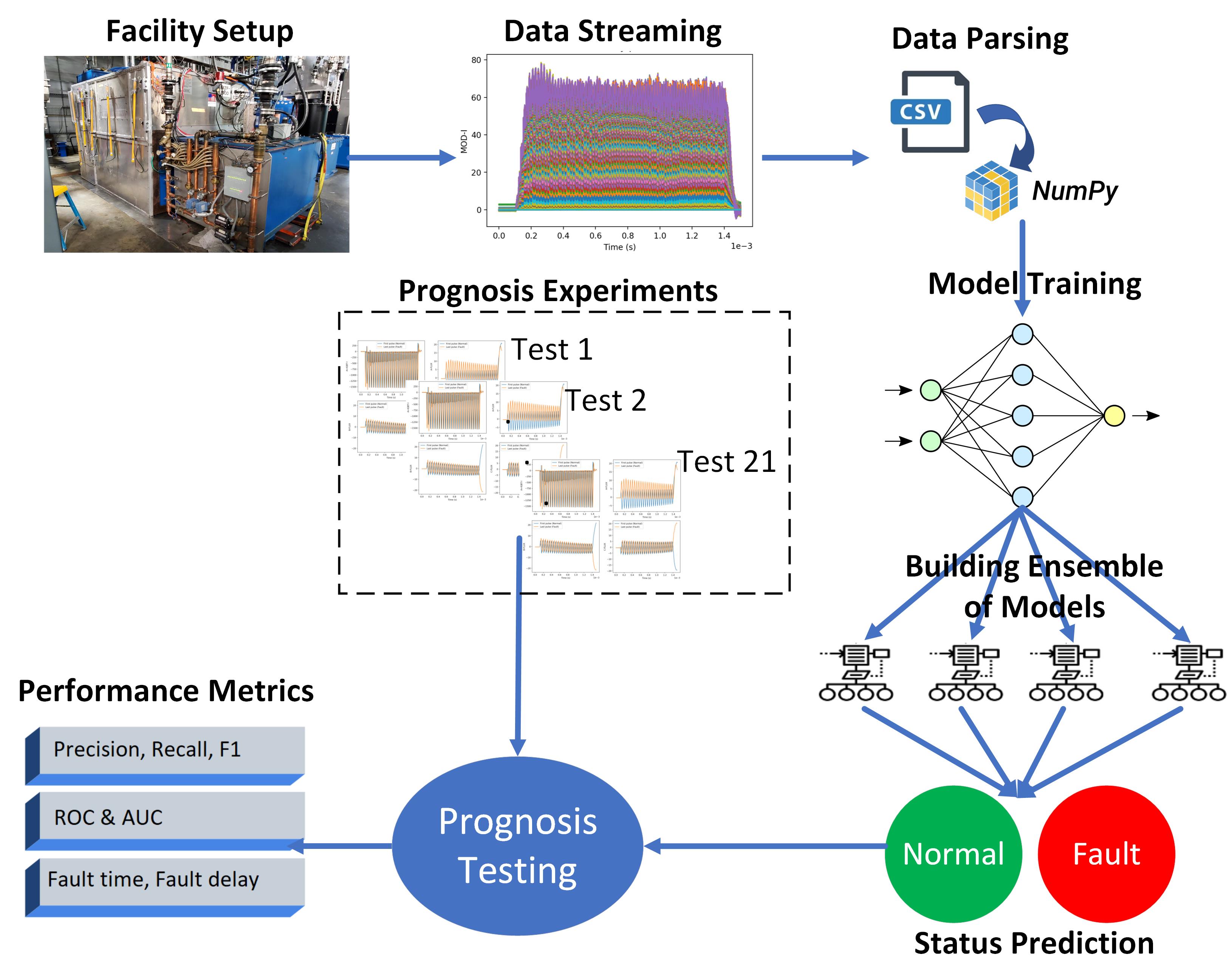}
   \caption{Methodology workflow}
   \label{fig:workflow}
 \end{figure*}

The methodology section involves description of data preparation, performing prognosis test scenarios, the proposed models, and performance metrics used to evaluate the models. Our methodology workflow is summarized in Figure \ref{fig:workflow}. The novelty of our methodology is that it lies in the intersection of robust experimental setup, quality data streaming, and machine learning modeling; each one of these parts is described next. 

\subsection{Training Data Preparation}
\label{sec:data}

Following data streaming for about 3 days in normal operating conditions, we collected large amounts of training data at a rate of a waveform file every 5 seconds. For faulty training data, these are a bit more challenging to collect in large amounts. However, we managed to collect about 5000 fault files that come from two main sources: (1) real faults in the RFTF and (2) data collected during HVCM tuning (fault-like data). The second source dominates the faulty training data given that real faults in the HVCM do not occur very frequently (i.e. weeks to few months). As described in the previous section, HVCM tuning is done manually by the operators following a HVCM startup. This process usually involves tweaking the HVCM settings to different values to optimize the waveform shapes. Fortunately, our data streaming is programmed to record pulses at the maximum saving rate (1 pulse per second) during the tuning process, and these tuning waveforms deviate from normal operating ranges. In normal conditions, HVCMs operate at full power, while the data during the tuning phase deviate from these full power conditions, which make them a great source of fault-like data (not real faults). To remove the class imbalance, we only kept 5000 normal waveform files that sufficiently cover normal conditions, to be consistent with the number of faulty files.  

After data streaming, the raw data are parsed from the CSV format and saved into a numpy array format with a 3D shape:
\begin{equation}
\label{eq:shape}
    shape=(N_{pulses} \times N_{times} \times N_{features})  
\end{equation}
where $N_{pulses}=10,000$ is the total number of pulses collected by the RFTF HVCM (both normal and faulty), $N_{times}$=3753 is the number of time steps for each pulse, and $N_{features}=12$ is the number of features or waveform types recorded for each pulse. Given the sampling rate is 400 ns, the time length of the pulse is approximately $3753 \times 400$ ns $\approx$ 1.5 ms. The 12 waveforms (features) were described in section \ref{sec:stream}.

\begin{figure}[!h]
   \centering
   \includegraphics*[width=0.8\textwidth]{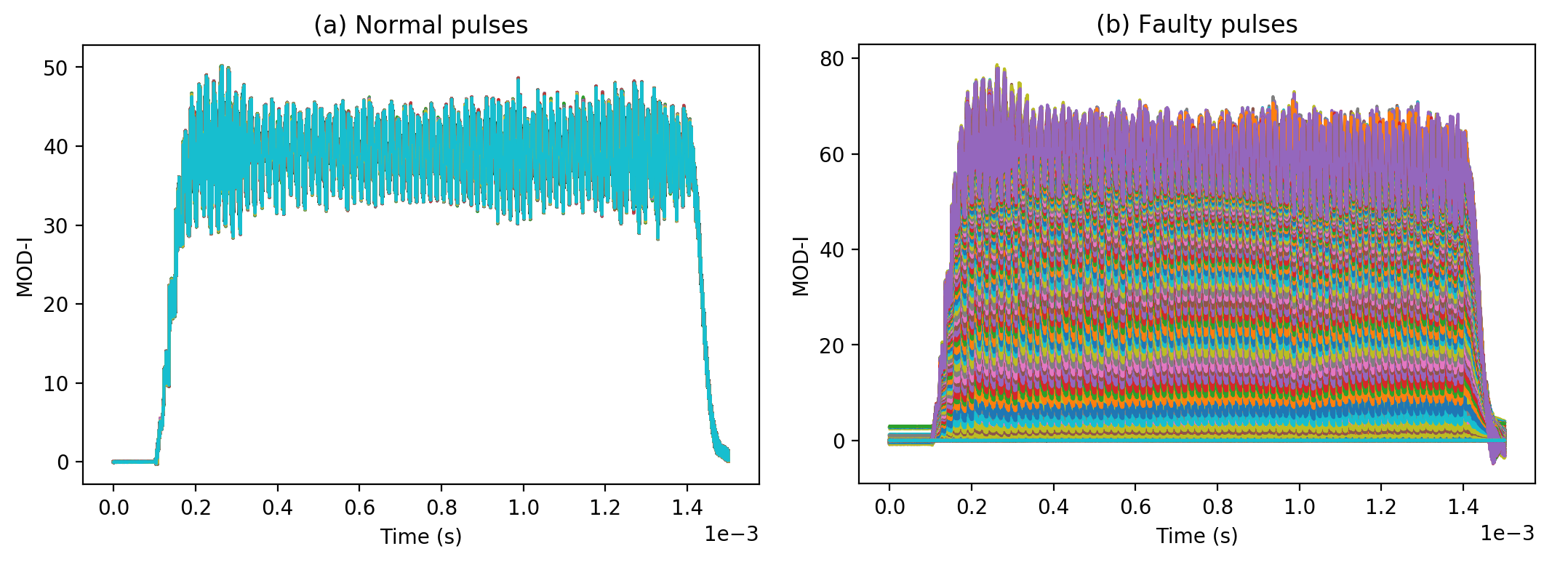}
   \caption{(a) 5000 normal modulator current (Mod-I) pulses, (b) 5000 faulty modulator current (Mod-I) pulses}
   \label{fig:train_data}
 \end{figure}

We plot the pulses/samples for the modulator current (Mod-I) in Figure \ref{fig:train_data}. The reader can easily see that the fault data, which consist primarily of tuning data, deviate from the normal Mod-I pulses that look like identical. It is worth mentioning again that tuning data are not real faults, but can expose machine learning models to conditions that deviate from the normal operating conditions. In both plots of Figure \ref{fig:train_data}, all 5000 normal and faulty pulses are shown, so no legend is provided. Other waveforms (A-Flux, CB-V, A+IGBT-I, etc.) exhibit similar behaviour when comparing normal to faulty data.

\subsection{Prognosis Tests}
\label{sec:prog}

\begin{figure*}[!h]
   \centering
   \includegraphics*[width=0.8\textwidth]{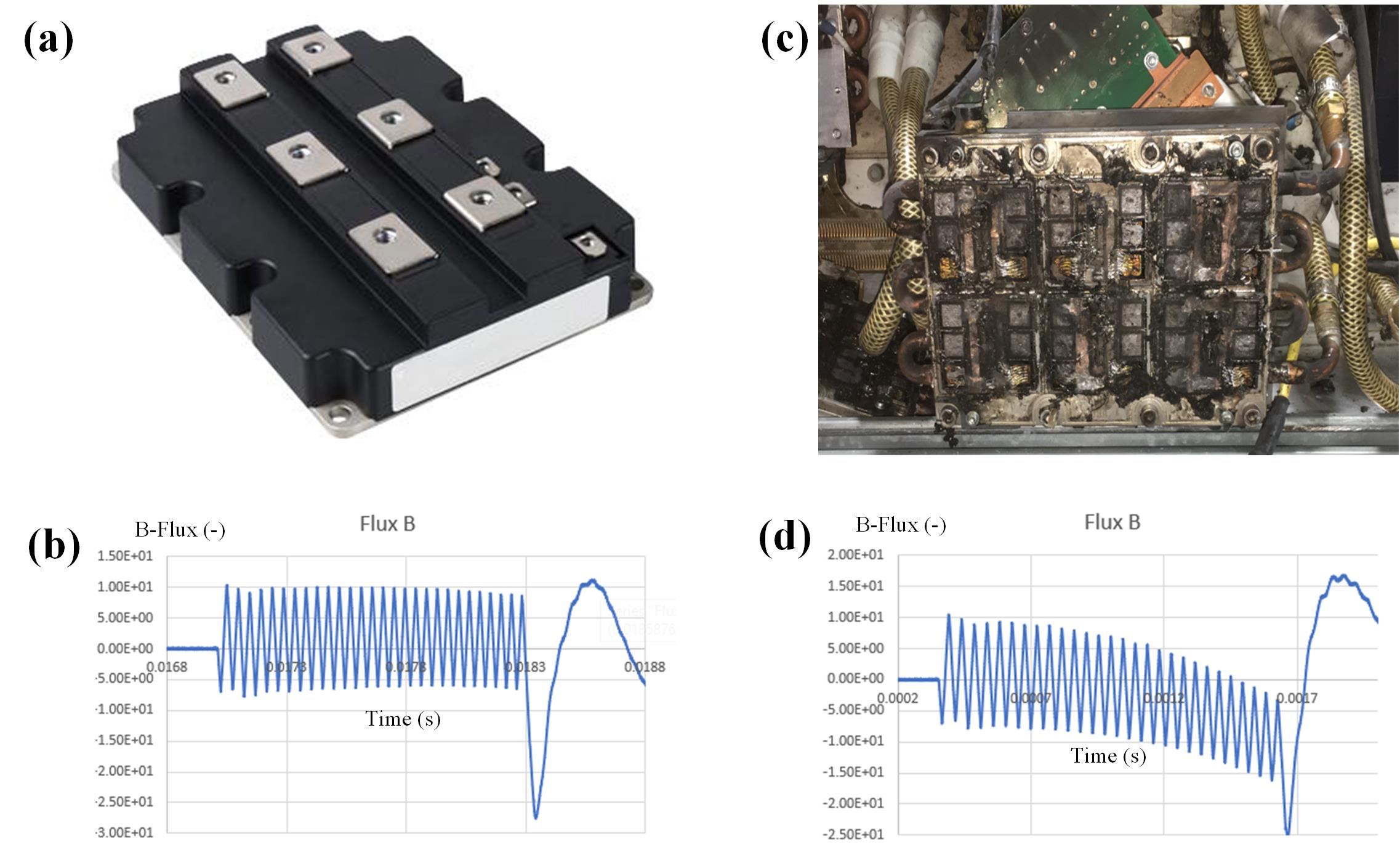}
   \caption{(a)-(b) Normal IGBT plate and its associated normal B-flux waveform. (c)-(d) Exploded IGBT plate and its associated faulty B-flux waveform (the pulse preceding the event)}
   \label{fig:igbt}
\end{figure*}

As described before, HVCMs continue to have failures which can range from mild faults that can be resolved by a system restart to catastrophic failures that can lead to IGBT explosion. Figure \ref{fig:igbt}(a)-(b) show a normal IGBT plate in the B phase along with its normal B-Flux waveform. In Figure \ref{fig:igbt}(c)-(d), the IGBT is exploded and the B-Flux waveform immediately before the event was recorded in Figure \ref{fig:igbt}(d), which shows a clear degradation and drooping in the flux signal. This fault event occurred during HVCM operation and was not planned by the authors.  

In this study, the authors have performed 21 independent experiments trying to simulate the common faults facing the HVCM like the one in Figure \ref{fig:igbt} by gradually inducing anomalous changes in the HVCM settings. These setting adjustments are sufficient to cause abnormality in the waveforms but not serious to cause a real fault. The idea is to create a continuous test in time where the machine starts in a normal condition and gradually moves to a fault condition, where the proposed models are assessed in their capabilities in detecting those changes as soon as possible. Each test involves the following steps:
\begin{enumerate}
\item The data streaming system in section \ref{sec:stream} is setup to save a waveform file every 7 seconds to allow the authors to make swift changes. 
\item The team starts every test by waiting about 3 minutes to collect normal waveforms using the normal settings. This period collects about 26 waveform files.  
\item The team then gradually induces changes in the RFTF settings and waits about a minute to collect waveform data under that change. 
\item The settings are changed by adjusting 9 knobs in Figures \ref{fig:controller1}-\ref{fig:controller2} to pre-established values determined by the team. These knobs fall under the categories of ``start pulse settings'' and ``flux compensation settings''. The changes can be in the form of increasing/decreasing start pulse width or increasing/decreasing flux compensation in the three phases. The adjustments can be in a single or multiple forms. Table \ref{tab:prog_tests} provides description of the plan and changes involved in each test. 
\item Each test finishes when the allowed (max, min) setting value is reached or when the system is in a serious condition that could lead to immediate failure. Therefore, the reader can expect the tests to have different time lengths as in Table \ref{tab:prog_tests}. 
\end{enumerate}

\begin{table*}[htbp]
\small
\centering
\caption{List of fault prognosis experiments conducted in the RFTF}
\begin{tabular}{lll}
\toprule
Test ID & Description*                                                                                      & Time (s) \\ \midrule
1        & A+ start pulse width   increases by 5\%/min                                                       & 868            \\
2        & B+ start pulse width   increases by 5\%/min                                                       & 707            \\
3        & C+* start pulse width   increases by 5\%/min                                                      & 616            \\
4        & A+ start pulse width   decreases by 5\%/min                                                       & 469            \\
5        & B+ start pulse width   decreases by 5\%/min                                                       & 581            \\
6        & C+* start pulse width   decreases by 5\%/min                                                      & 637            \\
7        & A-Flux compensation   increases by 25ns/min                                                        & 770            \\
8        & B-Flux compensation   increases by 25ns/min                                                        & 924            \\
9        & C-Flux compensation   increases by 25ns/min                                                        & 868            \\
10       & A-Flux compensation decreases   by 25ns/min                                                        & 812            \\
11       & B-Flux compensation decreases   by 25ns/min                                                        & 588            \\
12       & C-Flux compensation decreases   by 25ns/min                                                        & 693            \\
13       & A+ start pulse width   is set to 20\%, A-Flux compensation increases by 25ns/min                   & 924            \\
14       & A-*/B-*/C- start   pulse widths all decrease by 5\%/min                                           & 763            \\
15       & B+ start pulse width is   set to 100\%, B-Flux compensation decreases by 25ns/min                  & 707            \\
16       & A-Flux compensation   increases, B-Flux decreases, C-Flux increases by 25ns/min                    & 581            \\
17       & A-Flux compensation   decreases, B-Flux increases, C-Flux decreases by 25ns/min                    & 630            \\
18       & C+* start pulse width   is set to 90\%, C-Flux compensation increases by 25ns/min                  & 700            \\
19       & B-* start pulse width   is set to 50\%, B+ start pulse width increases by 5\%/min                 & 518            \\
20       & A+/A-* start pulse   widths are set to 20\%, A-Flux compensation decreases by 25ns/min             & 749            \\
21       & A-*/B-*/C- start   pulse widths are set to 40\%, A+/B+/C+* start pulse widths increase by 5\%/min & 532            \\ \bottomrule
\end{tabular}
\label{tab:prog_tests}%
\end{table*}

\begin{figure*}[!h]
   \centering
   \includegraphics*[width=\textwidth]{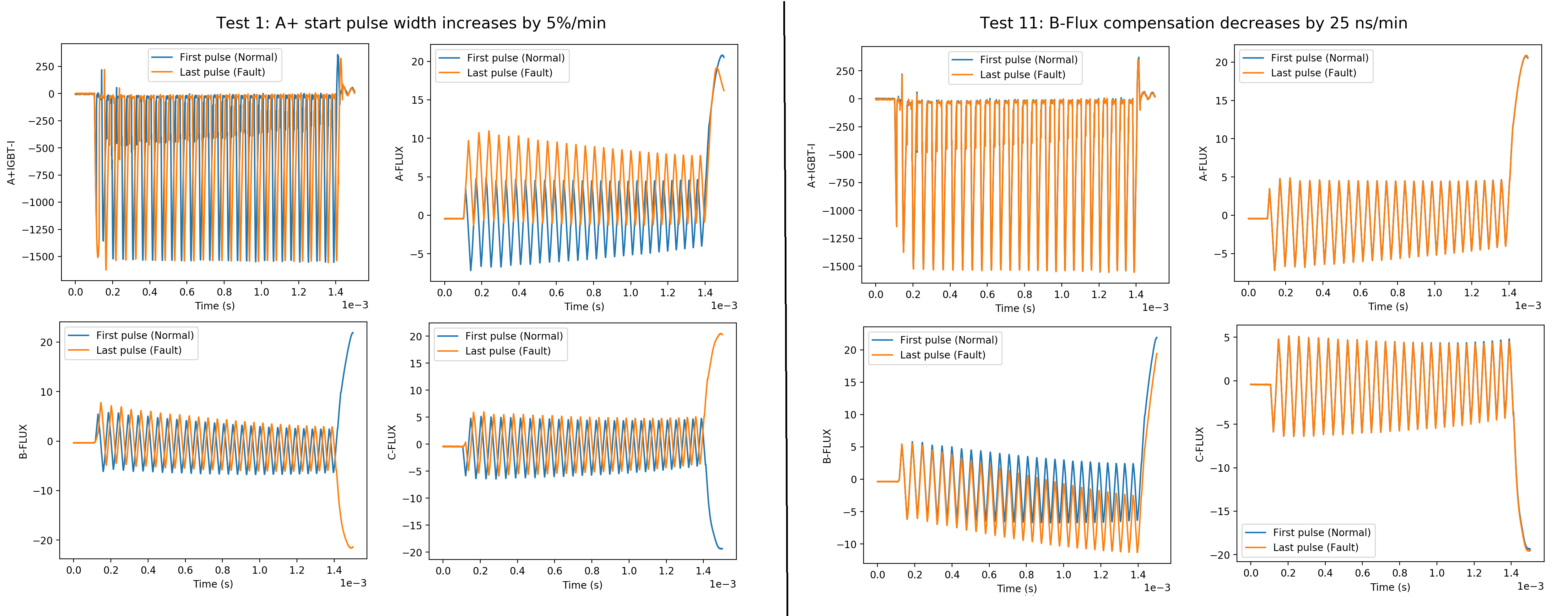}
   \caption{Test 1 (left) and Test 11 (right) results for four selected waveforms. The first pulse (before any change in settings) and the last pulse (after all changes are induced) are plotted}
   \label{fig:prog_waveforms}
\end{figure*}

We provide sample plots of four selected waveforms from Test 1 and Test 11 in Figure \ref{fig:prog_waveforms}, which involve increasing the A+ start pulse width and decreasing flux compensation in the B phase, respectively. In each subplot, the first pulse of the test (before any setting adjustment) and the last pulse (after all setting adjustments) are illustrated. These two tests along with the other 19 tests show interesting combinations of fault conditions that could be observed in a single or multiple waveforms. For example, the creeping of the fault precursors in Test 1 can be seen to cause an effect in the IGBT current and the fluxes of all phases (A, B, C). However, Test 11 reveals that decreasing the B-Flux compensation only affects the B-Flux waveform, while the fault precursors in other waveforms cannot be detected, which make these fault scenarios a bit tricky to detect.

For completeness, it is interesting mentioning that our original experiment list was planned to include 22 tests. The very last test which involved sudden changes in all 9 knobs/settings together indeed caused a real fault that happened in a fraction of a second after the 3 minutes normal run. Fortunately, the fault was a minor C+ driver fault, which was fixed by a system restart without causing any damage. The authors decided not to repeat that experiment to avoid causing a major trouble to the system, and we were satisfied by having 21/22 tests successful. The question to be answered next is: Can we develop a robust model that can detect fault precursors as soon as possible in most of the 21 prognosis tests?  

As the authors believe the current dataset can be useful for reproducibility of this study and for other researchers working in machine learning and fault prognosis areas, we shared all the data collected in this study. See the Data Availability section for more information.   

\subsection{Machine Learning Modeling}

\iffalse

\subsubsection{Feature selection}
\label{sec:features}

The authors have made comprehensive analysis of different combinations of the 12 waveform/features in this study (A+IGBT-I, A+*IGBT-I, ..., Mod-V, CB-V) to study their sensitivity and importance on the fault prognosis performance. Given the very high frequency of these signals (every $\sim$1.5ms), our criteria were dependent on minimizing signal pre-processing as much as we can to save that time for model prediction. Therefore, we explored options of using the raw waveforms as they come from the controller. In general, four different options of feature groups are highlighted in this paper:
\begin{enumerate}
    %\item Fitting the model with all 12 waveforms simultaneously. This option minimizes the number of models.
    \item Fitting a model based on the three flux waveforms (A-Flux, B-Flux, C-Flux).
    \item Fitting a model based on the all waveforms except the three flux waveforms (A+IGBT-I, A+*IGBT-I, B+IGBT-I, B+*IGBT-I, C+IGBT-I, C+*IGBT-I, Mod-I, Mod-V, CB-V).
    \item Fitting a model for each waveform independently. This option creates 12 independent models.
\end{enumerate}

It is worth mentioning that the authors have explored a much larger group of two-way and three-way combinations of waveforms without finding an interesting conclusion. The results of these four feature groups highlight most of the waveform impact on prognosis results. 
\fi

\subsubsection{Standalone Models}

The methods we select in this work belong to different categories including neural networks, ensemble of decision trees, and other classical methods. The convolutional neural network (CNN) classifier consists of Conv1D, max pooling, and fully-connected layers. The output is a binary prediction of the probability of a sample being normal or faulty pulse. The architecture of the CNN classifier is as follows:
\begin{enumerate}
    \item Reshape layer to convert 2D data shape compatible with other methods (random forests, bagging classifier, etc.) to 3D shape compatible with Conv1D layers.
    \item Conv1D layer with 32 filters, 6x6 kernel, ReLU activation, followed by max pooling of size 2x2.
    \item Conv1D layer with 32 filters, 4x4 kernel, ReLU activation, followed by max pooling of size 2x2.
    \item Conv1D layer with 16 filters, 3x3 kernel, ReLU activation, followed by max pooling of size 2x2.
    \item Conv1D layer with 8 filters, 2x2 kernel, ReLU activation, followed by max pooling of size 2x2.
    \item Flatten layer.
    \item Dense layer with 32 nodes and ReLU activation.
    \item Dense layer with 16 nodes and ReLU activation.
    \item Dense layer with 2 nodes and Softmax activation.
\end{enumerate}

The first reshaping layer is important to allow CNN to be trained as part of a large-scale ensemble that consists of other methods that support 2D data shape. 

Ensemble methods combine several models (e.g. decision trees) to produce better predictive performance than utilizing a single model. Ensemble methods can be broadly classified into bagging and boosting. Bagging builds a stronger model by reducing model ``variance'' through combining predictions of different models after using bootstrapping to create random subsets of the dataset with replacement, where these subsets are used to train their corresponding models. On the other hand, boosting builds a stronger model by reducing model ``bias''. In boosting, models are built sequentially where each subsequent model attempts to correct the errors of the previous model. Early models tend to be weak, and as boosting iteration continues, the weights of the models are adjusted to make correct predictions of the difficult samples that prior models failed to predict. In this study, we select four variants of ensemble methods based on bagging and boosting, all are based on randomized decision trees:
\begin{enumerate}
    \item Bagging Classifiers (BC) \cite{breiman1996bagging}: are ensemble estimators that fit base classifiers (e.g. randomized trees), each on random subsets of the original dataset, and then aggregate their predictions. The aggregation can be either by voting or by averaging their probabilistic prediction. Bagging methods are typically used to reduce the variance of the base estimators by introducing randomization into the procedure. In this work, we use BC with randomized trees and bootstrapping, where the subsets are drawn with replacement.
    \item Random forests (RF) \cite{breiman2001random}: are a special case of BC where the estimators are decision trees, while BC is a framework that can be used with other base estimators (e.g. support vectors, k-nearest neighbours). Nevertheless, RF still have more improvements in how the ensemble trees are built. For RF, when splitting each node during the construction of a tree, the best split is found either from all features or a random subset of the maximum number of features. RF achieve a reduced variance by combining diverse trees and use bootstrapping at the cost of a small increase in bias. We also use Gini impurity metric to measure the quality of a split.
    \item Extremely randomized trees (ET) \cite{geurts2006extremely}: are a modification of RF with two fundamental differences: (1) no bootstrapping (meaning ET samples without replacement) and (2) the nodes in ET are split on ``random'' splits rather than RF that use ``best'' splits, which could improve the performance.
    \item Adaboost (AB) \cite{freund1997decision}: The idea of AB is trying to fit a sequence of initial weak trees (i.e. random guessing) on modified versions of the dataset. The predictions from all trees are then combined using a weighted majority vote to produce the final prediction. The data modifications at each boosting iteration consist of applying weights to each of the training samples (starting by equal weights). Then the sample weights are modified and the algorithm is reapplied to the new weights in the next boosting iteration. After a certain number of iterations, the weights continue to adjust, where the algorithm focuses on the harder samples to predict to improve the overall performance. 
    \item Gradient Boosting (GB) \cite{friedman2001greedy}: GB is a generalization of AB to arbitrary differentiable loss functions compared to AB that uses the exponential loss function. For GB, binomial and multinomial deviance used in logistic regression is used as the loss function. GB also uses gradient descent to optimize the loss function.
\end{enumerate}

In all previous methods, the number of trees/estimators is the main tunable parameter.

Support vector machines (SVM) \cite{noble2006support, cervantes2020comprehensive} are well-known supervised learning methods used for classification and regression problems. Training is performed by defining separation hyperplanes, such as a line on a 2D surface or a plane in a 3D surface, between training samples to separate the data samples into distinct classes on the sides of the hyperplane. SVM can solve both linear and non-linear dataset problems by employing the kernel trick to transform the data to higher dimensions where the classes can be separable. Therefore, in this work, we use two variants of SVM: (1) linear SVM (LSVM) where the linear kernel is used and (2) non-linear SVM (RBF-SVM) where the radial basis function (RBF) kernel is used. 

\subsubsection{Hierarchical Voting Ensemble (VE)}
\label{sec:ve}

\begin{figure}[!h]
   \centering
   \includegraphics*[width=0.9\textwidth]{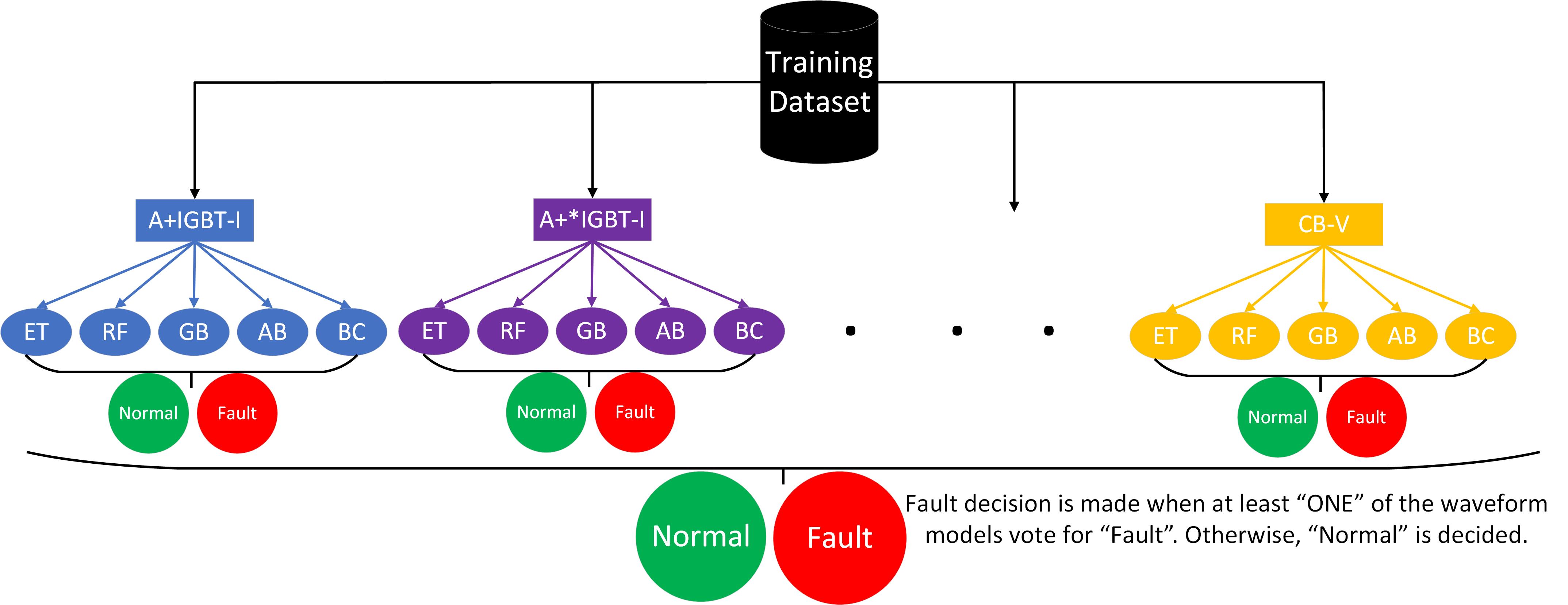}
   \caption{Hierarchical structure of the voting ensemble (VE) built in this work for fault prognosis}
   \label{fig:voting_classifier}
\end{figure}

Voting ensemble is the expression of ``democracy'' in the machine learning community and we adopt this approach here to minimize bias. The decision making process in deciding whether the signal is normal or faulty is made by taking the votes of all standalone models in the ensemble, and the decision is made by majority voting. The structure of the voting ensemble is shown in Figure \ref{fig:voting_classifier}, which can be described in the following steps:

\begin{enumerate}
    \item The training dataset is split into 12 subsets, each one highlights a single waveform data. 
    \item An ensemble model is trained by training 5 standalone models of RF, ET, GB, AB, and BC (i.e. these standalone models have shown the best performance when included in the ensemble).
    \item Each standalone model in the ensemble will vote for the signal type if it is faulty or normal, and the final decision is taken by majority voting. 
    \item After all 12 waveform models make their vote, the final decision whether the current condition is normal or faulty is decided when \textbf{at least one} of the waveform models votes for a fault condition. Otherwise, the decision is normal. 
\end{enumerate}

If the final decision is a fault, the operators will see the waveforms that have anomalies to help them diagnosing the fault source.

% \subsubsection{Hyperparameter Optimization}

% The voting ensemble model is tuned using parallel grid search to ensure optimal performance. The authors understand that excessive hyperparameter tuning could lead to a hyper-parameterized and rigid model performance that is crafted for the problem in hand. Therefore, given our ensemble has multiple models, we restrict the tuning to the following major hyperparameters that have the largest impact, while the rest are fixed to their default values:

% \begin{enumerate}
%     \item CNN: The $X$ value for the CNN model is selected from this grid ([8,16,24,32,48,64]), where $X$ expresses the number of filters and nodes for the hidden layers. 
%     \item GB: Number of estimators ([20,40,80,100,120]) and maximum number of features ([50,100,150,200]).
%     \item RF: Number of estimators ([20,40,80,100,120]) and maximum number of features ([50,100,150,200]).
%     \item BC: Number of estimators ([20,40,80,100,120]) and maximum number of features ([50,100,150,200]).
%     \item Adaboost: Number of estimators ([20,40,80,100,120]).
% \end{enumerate}

\subsection{Performance Metrics}
\label{sec:metrics}

We use standard metrics to evaluate the performance of our fault prognosis models in this study. Table \ref{tab:metrics} lists the classification metrics definition and their best and worst values. The four metrics are precision, recall, F1, and area under the curve (AUC), which is the area under the ROC curve (Receiver Operating Characteristic). The ROC curve is a standard diagram that illustrates the diagnostic ability of a binary classifier system as its discrimination threshold is varied. An AUC=0.5 indicates a random classifier without any classification capability. 

\begin{table*}[htbp]
\small
\centering
\caption{Classification performance metrics used to evaluate the proposed models}
\label{tab:metrics}%
\begin{threeparttable}
\begin{tabular}{lll}
\toprule
Metric & Formula* & Notes \\ \midrule
 Precision      &   $Precision=TP/(TP + FP)$   &   Best=1.0, Worst=0.0    \\
 Recall      &  $Recall = TP/(TP + FN)$        &   Best=1.0, Worst=0.0      \\
  F1     &   $F1 = 2TP/(2TP + FP + FN)$        &   Best=1.0, Worst=0.0     \\ 
  AUC     &  $AUC = \int_0^1 \text{TPR (FPR)} \ d\text{FPR}$    &    Best=1.0, Worst=0.5   \\ 
\bottomrule
\end{tabular}
 \begin{tablenotes}[para,flushleft]
  $^*$TP: True Positive, FP: False Positive, FN: False Negative, FPR: False Positive Rate, TPR: True Positive Rate.
  \end{tablenotes}
\end{threeparttable}
\end{table*}

In addition, we report the values for some prognosis metrics according to \cite{saxena2008metrics}. First, we report the time of detection of fault by the model ($t_F$), which we compare to the true fault time ($t_F^*$) in the experiment. The fault time $t_F$ is determined when the model predicts a fault event. If $t_F > t_F^*$, the delay in fault detection is determined by $\Delta t_F = t_F - t_F^* > 0$. Lastly, we report the fraction of the delay from the true total fault time as
\begin{equation}
    \rho = \frac{\Delta t_F}{t - t_F^*}\%,
\end{equation}
where $t$ is the total experiment time from Table \ref{tab:prog_tests}. The best value for $\rho$ is 0 (no delay) while the worst is 100\% (fault is missed). If $t_F < t_F^*$, which is the case when the model makes a fault prediction earlier than the true time, both $\Delta t_F$ and $\rho$ can have negative values. Given that the prognosis tests have different time lengths, the delay fraction provides more accurate representation of the fault delay.  

\section{Results and Discussions}
\label{sec:res}

\subsection{Analysis Settings}
\label{sec:hyper}

Following a set of preliminary tests, the following hyperparameters are utilized during the training process of all models:
\begin{enumerate}
    \item AB: Number of estimators is 40 and learning rate is 1.0. 
    \item BC, RF, ET: Number of estimators is 40.
    \item GB: Number of estimators is 40, max depth is 5, and learning rate is 0.1.
    \item CNN: Batch size of 64, 5 epochs, and Adam optimizer with $5 \times 10^{-4}$ learning rate. The loss function is the sparse categorical crossentropy.
    \item SVM: Regularization parameter has a value of 0.5 for LSVM and 1.0 for RBF-SVM
    \item VE: Inherits same hyperparameters of its members: BC, RF, ET, GB, and AB.
\end{enumerate}

Min-max scaling is applied to all waveforms to facilitate training. Given most methods support 2D data shape (except CNN), the second and third axes in Eq.\eqref{eq:shape} are flattened into a single axis, so the real number of features becomes $N_{times} \times N_{features} =  3753 \times 12 = 45036$. 

We have used Tensorflow/Keras with GPU support using CUDA and CuDNN libraries for the implementation of the CNN model. We also used Scikit-learn for the implementation of other machine learning models including our voting ensemble (VE). All training and analysis were conducted on a GPU cluster with 8 NVIDIA A100 SXM4 40GB GPUs available at the Spallation Neutron Source of the Oak Ridge National Laboratory.

\subsection{Training and Testing Results}

\begin{figure}[!h]
   \centering
   \includegraphics*[width=\textwidth]{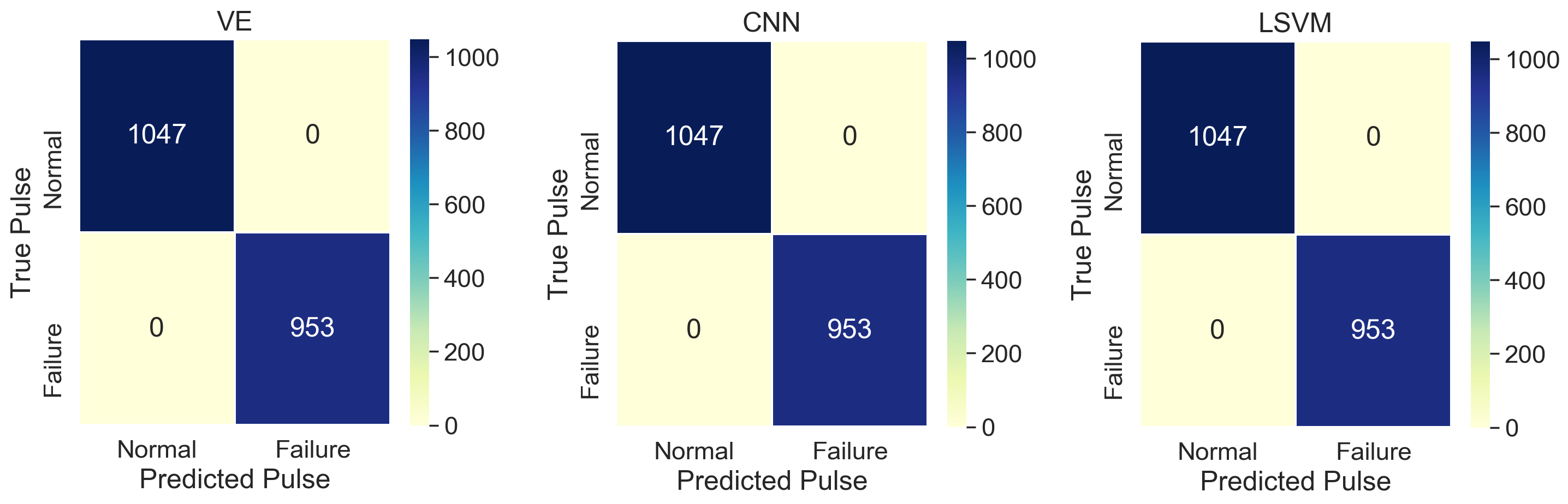}
   \caption{Confusion matrix based on the test set (2000 samples) for VE, CNN, LSVM. Other models (AB, BC, GB, RF, ET, RBF-SVM) show identical confusion matrix }
   \label{fig:confusion}
\end{figure}

In the training and testing phase, all models are trained with 8000 pulses/samples and tested with 2000 (i.e. 0.2 test split) using the data in Figure \ref{fig:train_data}. All 12 waveforms are included in the training process. The prognosis tests in Table \ref{tab:prog_tests} are not included in any sort during the training and testing phase. The results based on the test set indicate nothing but a superior and perfect performance by all 9 models, see Figure \ref{fig:confusion} which shows the confusion matrix for 3 selected models: VE, CNN, and LSVM. The confusion matrix shows that the three models did not make a single mistake in predicting the status of the 2000 test samples by having zero in the false positive and false negative entries. Other models (AB, BC, GB, RF, ET, RBF-SVM) show identical confusion matrix, but not shown for brevity. Accordingly, all 9 models have a perfect 1.0 for precision, recall, F1, and AUC, given the testing results. This perfect performance by all models reveals two interesting conclusions:
\begin{enumerate}
    \item The amount of training data (see Figure \ref{fig:train_data}) provided to all models is more than enough that makes all models performing very well at this stage, thanks to our facility setup and excellent data acquisition. 
    \item The values of the hyperparameters selected in section \ref{sec:hyper} for each model have no impact on the performance and are not causing any bias toward a certain model. 
\end{enumerate}

However, despite this amazing performance, the goal of this study is to predict the fault well ahead of time. Therefore, the question is, are all these perfect models going to generalize well for the 21 fault prognosis scenarios? 

\subsection{Prognosis Analysis}

To keep the paper concise, in this section, we focus the analysis on the VE model by presenting its corresponding results. The comparison with other models is provided in the second subsection. 

\iffalse

\subsubsection{Waveform Selection}
\label{sec:feature_res}

% Table generated by Excel2LaTeX from sheet 'waveform'
\begin{table}[htbp]
  \centering
  \small
  \caption{Effect of features/waveforms on prognosis performance (passed test is defined when the method achieves a delay fraction $\rho < 25\%$)}
  \begin{threeparttable}
    \begin{tabular}{lll}
    \toprule
    Waveform(s) Included in Training & Passed Test(s)* & Passed Test ID \\
    \midrule
    A+IGBT-I & 4     & 1, 4, 13, 20 \\
    A+*IGBT-I & 5     & 1, 4, 13, 20, 21 \\
    B+IGBT-I & 6     & 1, 4, 8, 13, 20, 21 \\
    B+*IGBT-I & 6     & 1, 4, 8, 13, 20, 21 \\
    C+IGBT-I & 3     & 4, 13, 20 \\
    C+*IGBT-I & 5     & 1, 4, 13, 20, 21 \\
    Mod-I & 5     & 1, 4, 13, 20, 21 \\
    A-Flux & 10    & 1, 4, 7, 10, 13, 14, 16, 17, 20, 21 \\
    B-Flux & 10    & 2, 4, 8, 13, 14, 15, 17, 19*, 20, 21 \\
    C-Flux & 12    & 1, 2, 3, 4, 9, 13, 14, 15, 16, 18, 20, 21 \\
    Mod-V & 3     & 1, 10, 21 \\
    CB-V  & 0     &  \\
    A-Flux, B-Flux, C-Flux & 7     & 4, 8, 13, 15, 19, 20, 21 \\
    Six IGBT-I (A+, A+*, B+, B+*, C+, C+*) & 6     & 1, 4, 8, 13, 20, 21 \\
    Mod-V, Mod-I, CB-V &       &  \\
    \bottomrule
    \end{tabular}%
  \label{tab:addlabel}%
  \begin{tablenotes}[para,flushleft]
  $^*$ Passed test is defined when the method achieves a delay fraction $\rho < 25\%$. 
  \end{tablenotes}
  \end{threeparttable}
\end{table}%

%table that compares number of tests passed given model

\fi

\subsubsection{Prognosis Results of VE}

The trained VE model is used to predict the fault timing of the 21 test scenarios using the hierarchy voting concept introduced in section \ref{sec:ve}. First, the classification metrics (precision, recall, F1, AUC) in distinguishing the normal from the faulty pulses in each test are listed in Table \ref{tab:class_ve}. The results show that the VE model still performs at a very high level by generalizing to the fault prognosis scenarios, obtaining almost perfect metrics for most of the test scenarios. For example, Tests 1-9, 13-14, 17-18, 20-21 show that the VE model is able to perfectly separate the pulses of the normal period (first 3 min) from the faulty pulses.  Other tests (10, 12, 15, 16, 19) also have excellent metrics. Test 11 is the only test with poor results as it seems VE is not able to classify the pulses of this test properly.

% Table generated by Excel2LaTeX from sheet 've_metrics'
\begin{table}[htbp]
  \centering
  \small
  \caption{Classification metrics when using the voting ensemble (VE) for the fault prognosis tests}
    \begin{tabular}{lllll}
    \toprule
    Test ID & Precision & Recall & F1    & AUC \\
    \midrule
    1     & 1.00  & 1.00  & 1.00  & 1.00 \\
    2     & 1.00  & 1.00  & 1.00  & 1.00 \\
    3     & 1.00  & 1.00  & 1.00  & 1.00 \\
    4     & 1.00  & 1.00  & 1.00  & 1.00 \\
    5     & 1.00  & 1.00  & 1.00  & 1.00 \\
    6     & 1.00  & 1.00  & 1.00  & 1.00 \\
    7     & 1.00  & 1.00  & 1.00  & 1.00 \\
    8     & 1.00  & 1.00  & 1.00  & 1.00 \\
    9     & 1.00  & 1.00  & 1.00  & 1.00 \\
    10    & 0.94  & 0.92  & 0.93  & 0.95 \\
    11    & 0.09  & 0.31  & 0.14  & 0.50 \\
    12    & 0.89  & 0.81  & 0.82  & 0.87 \\
    13    & 1.00  & 1.00  & 1.00  & 1.00 \\
    14    & 1.00  & 1.00  & 1.00  & 1.00 \\
    15    & 0.99  & 0.99  & 0.99  & 0.99 \\
    16    & 0.99  & 0.99  & 0.99  & 0.99 \\
    17    & 1.00  & 1.00  & 1.00  & 1.00 \\
    18    & 1.00  & 1.00  & 1.00  & 1.00 \\
    19    & 0.91  & 0.89  & 0.89  & 0.85 \\
    20    & 1.00  & 1.00  & 1.00  & 1.00 \\
    21    & 1.00  & 1.00  & 1.00  & 1.00 \\
    \bottomrule
    \end{tabular}%
  \label{tab:class_ve}%
\end{table}%

\begin{figure}[!h]
   \centering
   \includegraphics*[width=\textwidth]{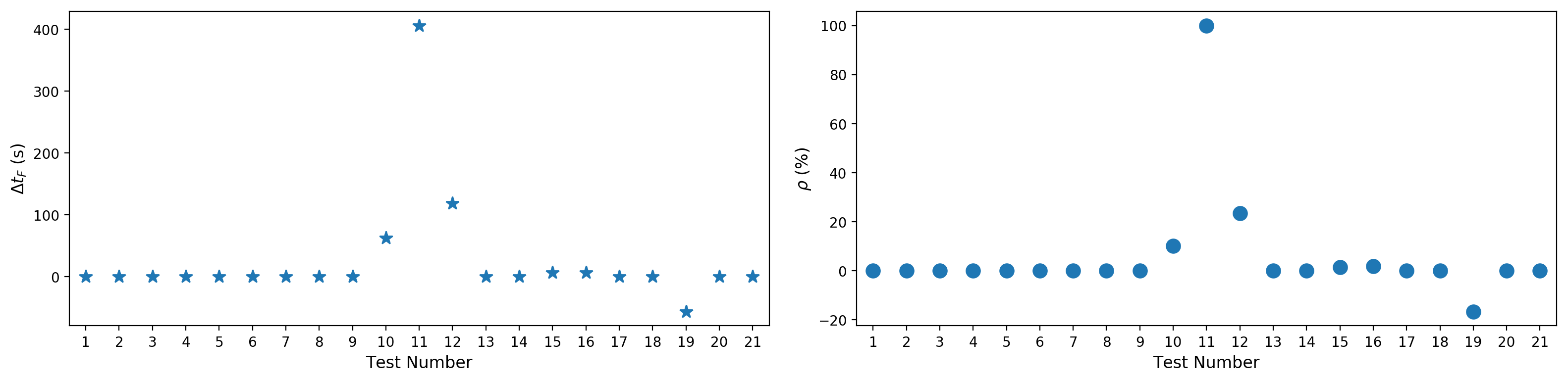}
   \caption{Prognosis performance of the voting ensemblle (VE) with fault delay time (left) and delay fraction (right). Test 11 is the only test missed by the model}
   \label{fig:prog_plot}
\end{figure}

Similarly, the prognosis metrics for the VE model are listed in Table \ref{tab:prog_ve}, which shows the detection time of the fault precursors ($t_F$), the true time of the fault precursors ($t_F^*$), time delay of the detection ($\Delta t_F$), and the fraction of the time delay from the true total fault precursor time ($\rho$). For better visualisation, the delay and delay fraction are plotted in Figure \ref{fig:prog_plot} for all 21 tests. In agreement with the results of Table \ref{tab:class_ve}, Figure \ref{fig:prog_plot} illustrates that VE is generalizing very well by detecting the fault precursors as soon as they actually appear in the system by achieving zero delay and delay fraction for Tests 1-9, 13-14, 17-18, 20-21. In addition, Tests 15, 16 show a very small delay fraction of 1\%, while Tests 10 and 12 show a larger but yet acceptable delay fractions of 10\% and 23\%, respectively. In Figure \ref{fig:prog_plot}, Tests 11 and 19 illustrate some distinct differences compared to the rest of the group. Test 11 is completely missed by the VE model as its delay fraction is the maximum 100\%, implying that the VE model was not predicting any fault signal after the first 3 min run. Test 11 is one of those tricky tests (see Figure \ref{fig:prog_waveforms}), where all but a single waveform remain identical to the normal period. For Test 11, all waveforms except the B-Flux remain almost identical the whole test time. And although the changes in the B-Flux waveform are visible, these were not enough for the VE model to detect them. Two main points should be discussed for Test 11:
\begin{enumerate}
    \item The total time of the test is about 588s (see Table \ref{tab:prog_tests}), which is on the shorter side of the time scale compared to the other tests. Also, the nature of the precursor being introduced (only the B-Flux compensation is adjusted) could imply that running Test 11 for longer times with more adjustments may improve the performance.
    \item As mentioned before, the authors avoided hyper-tuning of the proposed models to fit certain scenario(s), so the model can generalize well in the real world when it matters. Therefore, missing a single fault scenario and detecting the other 20 with a flexible model parameter set is already a major accomplishment. Nevertheless, we believe with careful tuning, the model can be crafted to detect Test 11 as well.  
\end{enumerate}

The second test with interesting results is Test 19, which shows negative time delay and delay fraction, implying that the model started to detect the fault precursors before they appear in the system. In this case 56 seconds earlier, which corresponds to -16.7\% delay fraction. This is an interesting observation, and after we explored the raw data of Test 19, we found that indeed the system was having a certain amount of noise in the waveform signals during the normal period, and that noise was significant enough for the VE model to detect them.     

% Table generated by Excel2LaTeX from sheet 've_progs'
\begin{table}[htbp]
  \centering
  \small
  \caption{Prognosis metrics when using the voting ensemble (VE) for the fault prognosis tests}
    \begin{tabular}{lllll}
    \toprule
    Test ID & $t_F^*$ (s)   & $t_F$ (s)    & $\Delta t_F$ (s) & $\rho (\%)$ \\
    \midrule
    1     & 182   & 182   & 0     & 0 \\
    2     & 182   & 182   & 0     & 0 \\
    3     & 182   & 182   & 0     & 0 \\
    4     & 182   & 182   & 0     & 0 \\
    5     & 182   & 182   & 0     & 0 \\
    6     & 182   & 182   & 0     & 0 \\
    7     & 182   & 182   & 0     & 0 \\
    8     & 182   & 182   & 0     & 0 \\
    9     & 182   & 182   & 0     & 0 \\
    10    & 182   & 245   & 63    & 10 \\
    11    & 182   & 588   & 406   & 100 \\
    12    & 182   & 301   & 119   & 23.3 \\
    13    & 182   & 182   & 0     & 0 \\
    14    & 182   & 182   & 0     & 0 \\
    15    & 182   & 189   & 7     & 1.3 \\
    16    & 182   & 189   & 7     & 1.8 \\
    17    & 182   & 182   & 0     & 0 \\
    18    & 182   & 182   & 0     & 0 \\
    19    & 182   & 126   & -56   & -16.7 \\
    20    & 182   & 182   & 0     & 0 \\
    21    & 182   & 182   & 0     & 0 \\
    \bottomrule
    \end{tabular}%
  \label{tab:prog_ve}%
\end{table}%

\begin{figure}[!h]
   \centering
   \includegraphics*[width=\textwidth]{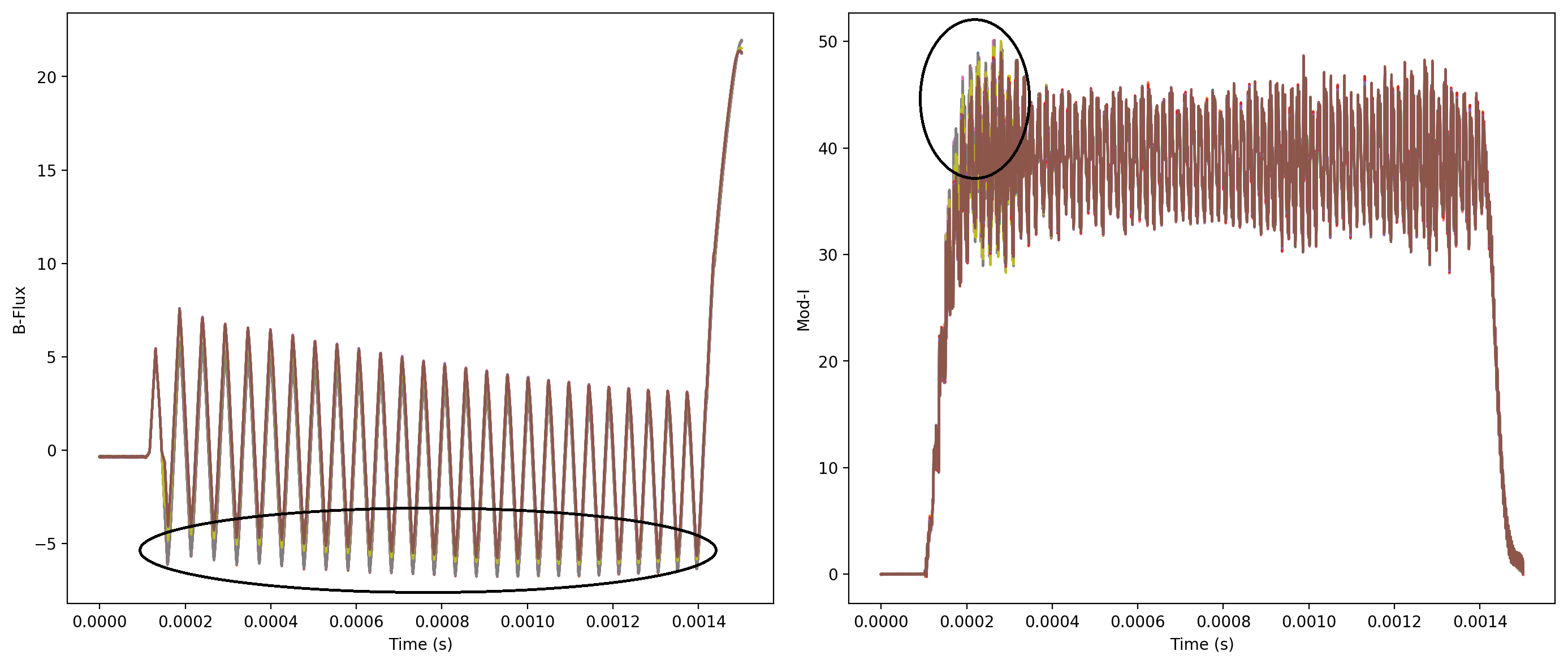}
   \caption{Plot of the B-Flux and Mod-I waveform pulses during the \textbf{second} and \textbf{third} minutes of Test 19}
   \label{fig:test19}
\end{figure}

To clarify more, Figure \ref{fig:test19} shows a plot of the B-Flux and Mod-I waveform pulses of Test 19. Only the pulses during the \textbf{second} and \textbf{third} minutes of the test are shown, which belong to the normal period when the VE model indicated fault precursors despite no precursors are introduced yet. The reader can clearly see in Figure \ref{fig:test19} that these normal pulses are not identical as desired, and some of them show clear deviation which can be seen more clearly in B-Flux and Mod-I waveforms. For example, the yellow and grey pulses obviously deviate from the rest of the group. This observation reveals an excellent proof that this model actually works and was able to detect those abnormal signals earlier in time even though they are not caused by a real fault precursor, but coming from system noise. While the authors are not certainly sure of the reason of why those abnormal signals show up in Test 19 in particular, the reason may be attributed to how these tests have been conducted. The authors have conducted these prognosis tests in a scheduled time frame where we usually leave 3-5 minutes before starting the next test. One possible reason could be that the machine was not able to restore its normal status following the several changes of Test 18, so small abnormalities remained and exacerbated after we started Test 19. As the reader can tell, these abnormalities are very difficult to detect by human unless a clear benchmark to normal signals is provided as in Figure \ref{fig:test19}. Test 19 results indeed show how these machine learning prognosis models can be really valuable in detecting such a subtle change in the waveforms. Now, after the normal period ends and real fault precursors start to appear, the VE model continues the excellent performance as can be easily told form the excellent classification metrics for Test 19 in Table \ref{tab:class_ve}. 

\subsubsection{Performance Comparison}

\begin{figure}[!h]
   \centering
   \includegraphics*[width=0.91\textwidth]{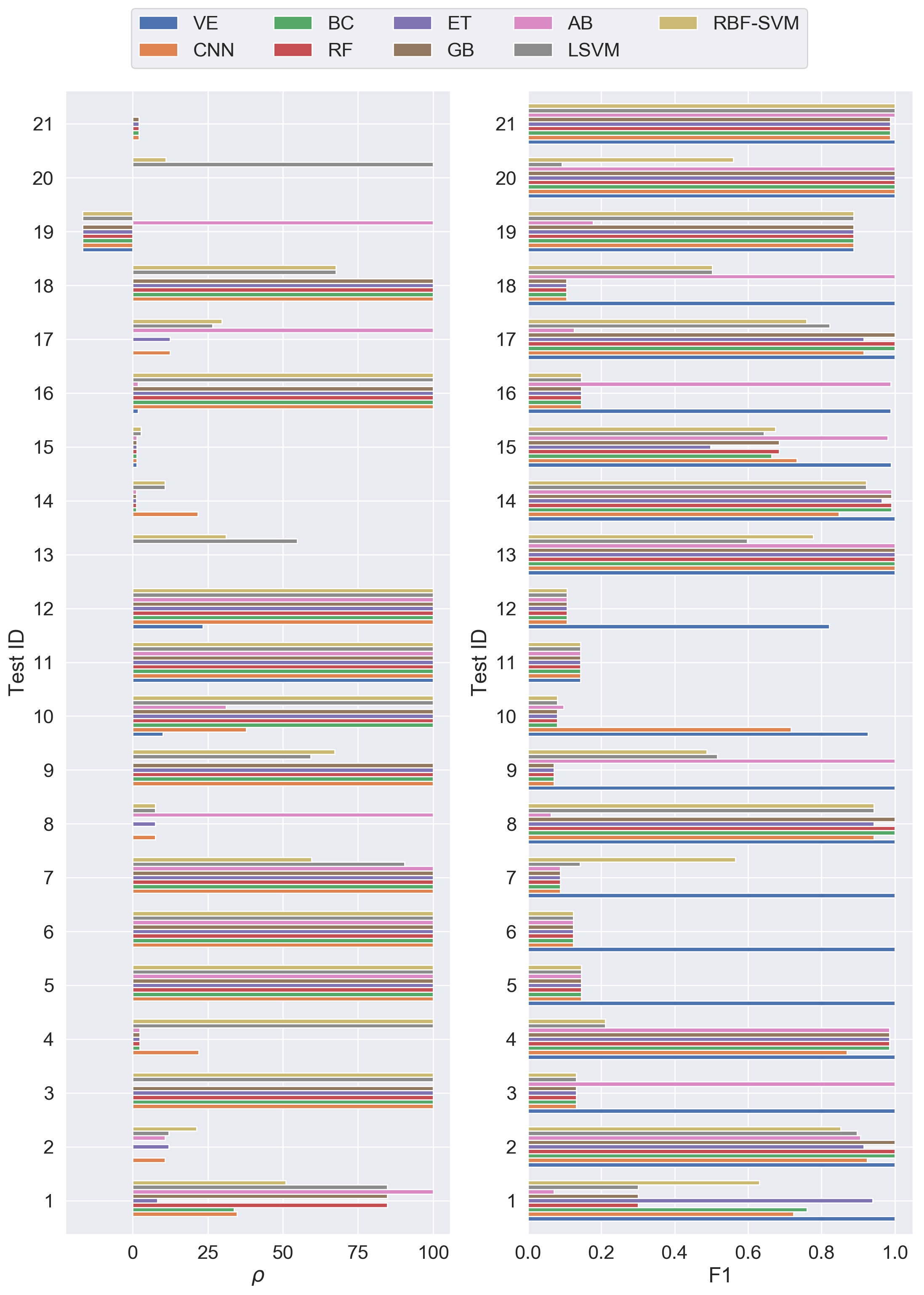}
   \caption{Comparison of the delay fraction ($\rho$) and F1 score for all methods and all tests}
   \label{fig:metrics_comp}
\end{figure}

All 9 trained models are now applied to the prognosis tests to evaluate their performance compared to the VE model, and the results are plotted in a bar chart in Figure \ref{fig:metrics_comp}. We only provided the delay fraction ($\rho$) and F1 score as representative prognosis and classification metrics, respectively. Due to the voluminous size of the results for all models and to maintain a concise article, we only reported these two metrics in this paper. \textit{However, all metrics for all 9 models are provided in a spreadsheet in the supplementary materials of this article.} 

By looking at the results, it is clear that the perfect models in the training/testing phase are no longer that good coming into the prognosis phase. The models do not seem to provide a satisfactory performance compared to VE in most of the tests. For example, Tests 5-7, the models other than VE have missed the fault precursors with a large delay (sometimes 100\%) and have provided a poor F1 score, while VE shows the opposite. And this is the same story for most of the other tests, however, with some tests showing variability in performance between the models. For example, Test 1 shows ET as the best method after VE, Test 2 shows that BC, RF, GB are as good as VE, Test 3 shows AB and VE to be the best, and so on. The major observations to be discussed from Figure \ref{fig:metrics_comp}:
\begin{itemize}
    \item All models including VE failed to detect Test 11, obtaining comparable metrics. This could imply that the precursors of Test 11 are hard to detect. 
    \item All models except AB show a negative delay fraction for Test 19, implying that the noise precursors were detectable by most models. This agreement reinforces our observations for VE. AB demonstrated a poor performance, missing whole Test 19. 
    \item Test 15 shows that all models have comparable delay fractions, but by looking at their F1 score, it implies something else. Although all models were able to detect the precursors early enough, some models started to make wrong predictions afterwards, tagging faulty pulses as normal. For example, VE and AB have very good F1 and $\rho$ scores, but ET has a good $\rho$ but a mediocre F1 score of less than 0.6. This shows the value of looking at different metrics to fully assess model performance.   
\end{itemize}

To provide a comprehensive and concise comparison between all models based on these metrics, we counted the number of tests being passed by each model. The passing condition is to achieve a delay fraction $\rho < 25\%$ and $F1 > 0.8$. While these thresholds are arbitrary for evaluation, a 25\% delay still provides the operator with enough time to take an action. The summary is provided in Table \ref{tab:summary}, which shows how these models are ranked. Obviously, the voting classifier (VE) excels by far passing 20 out of 21 tests, yielding an impressive success rate of 95\%. The rest of the models are ranked next with comparable performances, with AB and ET coming next with 52\% and 48\% success rates, respectively. The CNN, BC, RF, GB have similar overall performance, while the two SVM models (LSVM and RBF-SVM) show the worst performance without any advantage of adding the RFB kernel over the linear kernel. 

% Table generated by Excel2LaTeX from sheet 'summary'
\begin{table}[htbp]
  \centering
  \caption{Comprehensive summary of the methods passing the prognosis tests with: $\rho < 25\%$ and $F1 > 0.8$ and their fraction of the total (success rate)}
   \begin{threeparttable}
    \begin{tabular}{lrr}
    \toprule
    Method & \multicolumn{1}{l}{Passed Tests} & \multicolumn{1}{l}{Success Rate (\%)*} \\
    \midrule
    VE    & 20    & 95.2 \\
    CNN   & 9     & 42.9 \\
    BC    & 9     & 42.9 \\
    RF    & 9     & 42.9 \\
    ET    & 10    & 47.6 \\
    GB    & 9     & 42.9 \\
    AB    & 11    & 52.4 \\
    LSVM  & 5     & 23.8 \\
    RBF-SVM & 5     & 23.8 \\
    \bottomrule
    \end{tabular}%
   \begin{tablenotes}[para,flushleft]
  $^*$ Success rate is the fraction of the passed tests from the total number of tests (21). 
  \end{tablenotes}
  \end{threeparttable}
  \label{tab:summary}%
\end{table}%

The proposed models in this work encompass a wide range of machine learning methods from ensembles to neural networks to classical methods like SVM. The VE model, the main hierarchical ensemble based on the voting concept, illustrated to be the best. The VE model is an ensemble of multiple layers, each consists of an ensemble as shown in Figure \ref{fig:voting_classifier}. The first layer features different models, each voting on the status of its corresponding waveform. The second layer involves using five different sub-models within each waveform model to vote on the status of that waveform. In this context, we used models based on random forests, extremely randomized trees, bagging classifiers, adaboost, and gradient boosting. The third layer occurs within each sub-model (RF, BC, etc.), where each sub-model consists of number of estimators (i.e. 40) voting for each sub-model. This advanced hierarchy provides a strong diversity in the decision making process which is the main reason for the excellent performance of VE. The second advantage of this approach is that it significantly reduces the sensitivity of the hyperparameters of each model/estimator given the decision is made by more than 100 models rather than a single model. 

The remaining models show that standalone ensembles (GB, BC, etc.) can be as good as neural network models (CNN) for prognosis, while classical models such as SVM seem to be less powerful. For completeness, we also tested k-nearest neighbours (KNN) and feedforward neural networks, with both showing a poor prognosis performance that are not worth to be reported here. This is aside from the fact that KNN was very slow to train and to make predictions. Also, we explored including the CNN, LSVM, and RSVM into the VE ensemble without noticing major improvements, albeit that the SVM models made VE even weaker. 

Aside from the machine learning part, we should reiterate on the significant value of the experimental setup used in this study, which we believe is as important as the machine learning part. The experimental part involves significant efforts by electronics and control engineers to facilitate data streaming and prognosis tests, which were vital to obtain quality and adequate quantity of data to empower machine learning. The RFTF facility can be used in future computational and machine learning studies to explore these techniques in other parts of the accelerator.

The results of this study also reveal that the classical machine learning approach of splitting data into training and test sets is not always guaranteed to confirm the model performance. In this work, 9 models show identical and perfect performance on the test set, but once they are introduced to new prognosis data, the performance of 8 out of 9 decreased at least by half. This practice reveals the value of implementing the trained models and expose them to new data to confirm their ability to generalize.  

In terms of computing time, the training expenses are quite comparable for most models except for VE, which is more expensive than others as it involves many more models to train. However, given that training is done offline, and the trained model is what is being used on the system, the model prediction time is more important. Similarly, we expect the VE model to be also slower as it counts votes from many sub-models to make a prediction. The tests reveal that RBF-SVM is the slowest model among all taking prediction time of 28 ms/sample, most likely due to the non-linear transformation. Next is the VE model taking 6 ms/sample, then LSVM with 3 ms/sample, CNN with 2 ms/sample, while the rest of the models (GB, BC, GB, RF, ET) take in average between 1.5-1.8 ms/sample. Given we are streaming waveforms at a rate of 3-5 s, the prediction time of all models is much lower.   

In this work, we accomplished two main goals that were limitations of our previous effort \cite{radaideh2022time}: (1) resolving data limitations by upgrading the controller and the data acquisition system of the HVCM, (2) demonstrating that under near-continuous data streams, machine learning can be effectively used for fault prognosis to detect the fault precursors well ahead of time. In the previous paper \cite{radaideh2022time}, which highlights the HVCM powering the RFQ section, the paper demonstrates a promising potential for fault detection, but with a very limited time scale (about 1.5 ms before the fault happens), which is a time that only allows for a quick system shutdown. This approach may prevent the fault from happening and reduce the damage to the HVCM (i.e. through preventive maintenance), but does not resolve the downtime issue of the SNS. The results of this work, which are based on the RFTF facility simulating the SNS conditions, extend these methods to allow for fault prognosis and predictive maintenance, giving the operators sufficient time to either re-tune the modulator, skipping the warning if it is only a noisy signal, or shutting down the system if the issue is serious.  

\section{Conclusions}
\label{sec:conc}

In this work, variety of machine learning methods are tested in performing fault prognosis and detection in particle accelerator power electronics to reduce their catastrophic failures and improve the particle accelerator reliability. The study highlights the spallation neutron source (SNS) and the high voltage converter modulators (HVCM) that power the klystrons and accelerating cavities. An advanced experimental setup featuring a radio-frequency test facility with operating conditions similar to the SNS is used to stream waveform data in much higher rates than what was achieved before. The authors have conducted 21 prognosis experiments mimicking the fault conditions that occurred in the HVCM in the past without causing a real fault, where machine learning models are tested in discovering these fault precursors as soon as they are introduced in the system. Variety of techniques including ensemble trees, convolutional neural networks, support vector machines, and hierarchical voting ensembles are trained and tested in this study. Although all models have shown a perfect and identical performance during the training and testing phase, the performance of most models has decreased in the prognosis phase once they got exposed to real-world data that feature the 21 experiments. Nevertheless, the hierarchical voting ensemble maintains a distinguished performance in early detection of the fault precursors in 20 out of 21 tests, followed by adaboost and extremely randomized trees that detected 11 tests and 10 tests, respectively. The support vector machine models provided the worst performance detecting only 5 tests. The performance of the other models was in between.

Overall, the results of this study lead to several conclusions that for a successful implementation of machine learning in the SNS or particle accelerator power systems (e.g. HVCM), the data acquisition system should be improved to be more advanced and capable for continuous streaming and handling of big data to feed to the machine learning models. The machine learning models are better to be diverse and based on ensemble concepts to reduce bias and hyperparameter sensitivity. Given the authors have shared all experimental data used in this work, future work ideas would focus on improving the models architecture and their generalizing abilities to allow for accurate prognosis and predictive maintenance.

\section*{Acknowledgment}

The authors are grateful for support from the Neutron Sciences Directorate at ORNL in the investigation of this work. This work was supported by the DOE Office of Science under grant DE-SC0009915 (Office of Basic Energy Sciences, Scientific User Facilities program). This research used resources of the Compute and Data Environment for Science (CADES) at the Oak Ridge National Laboratory, which is supported by the Office of Science of the U.S. Department of Energy under Contract No. DE-AC05-00OR22725. The authors would like to thank Willem Blokland from the spallation neutron source for his useful comments and feedback on this work. 

Notice: This manuscript has been authored by UT-Battelle, LLC, under contract DE-AC05-00OR22725 with the US Department of Energy (DOE). The US government retains and the publisher, by accepting the article for publication, acknowledges that the US government retains a nonexclusive, paid-up, irrevocable, worldwide license to publish or reproduce the published form of this manuscript, or allow others to do so, for US government purposes. DOE will provide public access to these results of federally sponsored research in accordance with the DOE Public Access Plan (http://energy.gov/downloads/doe-public-access-plan). 

\section*{Data Availability}

The data will be shared in a Mendeley repository following the peer-review of this work to ensure the authors' credit is fully appreciated. The data will include the training data as well as the prognosis tests. Also, a companion data article is prepared to describe how to utilize the dataset. 

%The reviewers can access this companion data paper in the supplementary materials of this submission. The data paper does not affect or add to the flow of this paper, and it is only provided as an additional reference.
%Temporary link to the data:} \href{https://www.dropbox.com/sh/lp0zbrpyv7nfgwg/AAAtmY9dFkqj0kyXk6n6Z_jfa?dl=0}{[LINK]}

\section*{CRediT Author Statement}

\noindent \textbf{Majdi I. Radaideh}:  Conceptualization, Methodology, Software, Validation, Investigation, Data curation, Visualisation, Formal analysis, Writing - Original Draft. \\
\textbf{Chris Pappas}: Conceptualization, Data Curation, Software, Writing – Review and Edit. \\
\textbf{Mark Wezensky}: Data Curation, Software, Validation, Writing – Review and Edit. \\
\textbf{Pradeep Ramuhalli}: Conceptualization, Methodology, Supervision, Writing – Review and Edit. \\
\textbf{Sarah Cousineau}: Conceptualization, Funding acquisition, Project Administration, Writing – Review and Edit.

%\section*{References}

\bibliographystyle{elsarticle-num}
%\bibliographystyle{apa}
%\biboptions{authoryear}
\setlength{\bibsep}{0pt plus 0.3ex}
{
\footnotesize \bibliography{references}}

\end{document}